\newif\ifdraft\drafttrue 
\documentclass[letterpaper,twocolumn,10pt]{article}
\usepackage{usenix2019_v3}
\ifdraft
\usepackage[switch]{lineno}
\fi

\usepackage[utf8]{inputenc}
\usepackage[english]{babel}
\usepackage[numbers,sort]{natbib}

\usepackage{epsfig}
\usepackage{endnotes}

\usepackage{booktabs}   
\usepackage{subcaption} 
\usepackage{microtype}
\usepackage{xspace}
\usepackage{url}
\usepackage{tabularx,booktabs}
\usepackage{pbox}
\usepackage{xcolor}
\usepackage{listings}
\usepackage{float}
\usepackage{verbatim}
\usepackage{wrapfig}
\usepackage{multirow}
\usepackage{makecell}
\usepackage{etoolbox}
\usepackage{tikz}

\hypersetup{%
  colorlinks=true,
  citecolor=black,
  linkcolor=black,
  urlcolor=black
}

\microtypecontext{spacing=nonfrench}
\microtypesetup{
  final,
  tracking=true,
  kerning=true,
  spacing=true,
  factor=1100,
  stretch=20,
  shrink=20
}
\SetTracking{encoding={*}, shape=sc}{0} 

\newcommand{\asmjs}{\texttt{asm.js}\xspace}
\newcommand{\wasm}{WebAssembly\xspace}
\newcommand{\browsix}{\textsc{Browsix}\xspace}
\newcommand{\Browsix}{\textsc{Browsix}\xspace}
\newcommand{\browsixwasm}{\textsc{Browsix-Wasm}\xspace}
\newcommand{\browsixWasm}{\textsc{Browsix-Wasm}\xspace}
\newcommand{\Browsixwasm}{\textsc{Browsix-Wasm}\xspace}
\newcommand{\BrowsixWasm}{\textsc{Browsix-Wasm}\xspace}
\newcommand{\BrowsixSpec}{\textsc{Browsix-SPEC}\xspace}
\newcommand{\BrowsixSPEC}{\textsc{Browsix-SPEC}\xspace}
\newcommand{\BrowserFS}{\textsc{BrowserFS}\xspace}

\lstdefinelanguage{myC}{
  language=C,
  columns=fullflexible,
  keepspaces=true,
  keywordstyle=\color{blue}\ttfamily,
  stringstyle=\color{red}\ttfamily,
  commentstyle=\color{green}\ttfamily,
  morecomment=[l][\color{magenta}]{\#}
}

\lstdefinelanguage{x8664intel}{
  keywords={xor,mov,imul,add,lea,mov,imul,add,cmp,pop,ret,
            jne, xor, nop, cmp, jz, jmp, cmp, ret, jnz},
  keywordstyle={\color{red}},
  comment=[l][\color{magenta}]{\#},
  columns=fullflexible,
  keepspaces=true,
  mathescape=true,
  emph={eax, ebx, ecx, edx, rax, rbx, rcx, rdx, r8d, r9d, r10d, r12d,
        r13d, r14d, r15d, r8, r9, r10, r11, r12, r13, r14, r15, rsi, rdi,
        rsp, rsb, esp, esi, edi, esb, r11d},
  emphstyle={\color{blue}}
}

\lstdefinestyle{figurestyle}{
  basicstyle=\fontsize{10}{12}\ttfamily,
  numbers=left,
  backgroundcolor=\color{yellow!10},
  numberstyle=\fontsize{5}{5}\rmfamily,
  numbersep=0.4em,
  stepnumber=1,
  firstnumber=1}

\lstdefinestyle{normalstyle}{
  basicstyle=\small\ttfamily\fontsize{7}{7},
  numbers=none}

\newtoggle{infig}
\togglefalse{infig}
\AtBeginEnvironment{figure}{\toggletrue{infig}}
\AtEndEnvironment{figure}{\togglefalse{infig}}
\AtBeginEnvironment{subfigure}{\toggletrue{infig}}
\AtEndEnvironment{subfigure}{\togglefalse{infig}}
\AtBeginEnvironment{figure*}{\toggletrue{infig}}
\AtEndEnvironment{figure*}{\togglefalse{infig}}

\AtBeginEnvironment{lstlisting}{
  \iftoggle{infig}{
    \lstset{style=figurestyle}
  }{
    \lstset{style=normalstyle}}}

  \pretocmd{\lstinputlisting}{%
  \iftoggle{infig}{%
    \lstset{style=figurestyle}%
  }{%
    \lstset{style=normalstyle}}}{}{}

\begin{document}

\date{}

\title{\Large \bf Not So Fast: \\ Analyzing the Performance of WebAssembly vs. Native Code}

\author{\rm
{Abhinav Jangda, Bobby Powers, Emery D. Berger, and Arjun Guha}\\
University of Massachusetts Amherst}

\maketitle
\pagenumbering{gobble}


\begin{abstract}
All major web browsers now support WebAssembly, a low-level bytecode
intended to serve as a compilation target for code written in
languages like C and C++. A key goal of WebAssembly is performance
parity with native code; previous work reports near parity, with many applications
compiled to WebAssembly running on average $10\%$ slower than native
code. However, this evaluation was limited to a suite of scientific
kernels, each consisting of roughly 100 lines of code. Running more
substantial applications was not possible because compiling code to
WebAssembly is only part of the puzzle: standard Unix APIs are not
available in the web browser environment. To address this challenge,
we build \browsixwasm, a significant extension
to \browsix~\cite{powers:2017:browsix} that, for the first time, makes
it possible to run unmodified WebAssembly-compiled Unix applications
directly inside the browser. We then use \BrowsixWasm to conduct the
first large-scale evaluation of the performance of WebAssembly
vs. native. Across the SPEC CPU suite of benchmarks, we find a
substantial performance gap: applications compiled to WebAssembly run
slower by an average of 45\% (Firefox) to 55\% (Chrome), with peak
slowdowns of $2.08\times$ (Firefox) and $2.5\times$ (Chrome).  We
identify the causes of this performance degradation, some of which are
due to missing optimizations and code generation issues, while others
are inherent to the WebAssembly platform.

\end{abstract}

\section{Introduction}\label{sec:introduction}

Web browsers have become the most popular platform for running user-facing
applications, and until recently, JavaScript was the only programming language
 supported by all major web browsers. Beyond its many quirks
and pitfalls from the perspective of programming language design, JavaScript is
also notoriously difficult to compile efficiently~\cite{bebenita:spur,gal:tracemonkey,selakovic:js-optimizations,richards:js-behavior}. Applications
written in or compiled to JavaScript typically run much slower than
their native counterparts. To address this situation, a group of
browser vendors jointly developed \emph{WebAssembly}.

WebAssembly is a low-level, statically typed language that does not require
garbage collection, and supports interoperability with JavaScript. The goal of
\wasm{} is to serve as a universal compiler target that can run in a
browser~\cite{elliott:2015wasm,eich:2015wasm,haas:2017:webassembly}.\footnote{The
\wasm{} standard is undergoing active development, with ongoing efforts to
extend \wasm{} with features ranging from SIMD primitives and threading to tail
calls and garbage collection. This paper focuses on the initial and stable
version of \wasm{}~\cite{haas:2017:webassembly}, which is supported by all
major browsers.} Towards this end, \wasm{} is designed to be fast to compile
and run, to be portable across browsers and architectures, and to provide
formal guarantees of type and memory safety. Prior attempts at
running code at native speed in the
browser~\cite{activeX,pnacl,nacl-and-pnacl,asmjs}, which we discuss in related
work, do not satisfy all of these criteria.

\wasm{} is now supported by all major
browsers~\cite{wagner:2016wasm,webassembly} and has been swiftly adopted by
several programming languages. There are now backends for C, C++, C\#, Go, and
Rust~\cite{emscripten,musiol:2016gopherjs,rust-wasm,blazor} that target
\wasm{}. A curated list currently includes more than a dozen
others~\cite{curated-wasm-list}. Today, code written in these languages can be
safely executed in browser sandboxes across any modern device once compiled to
WebAssembly.

\begin{figure}
  \includegraphics[width=\linewidth]{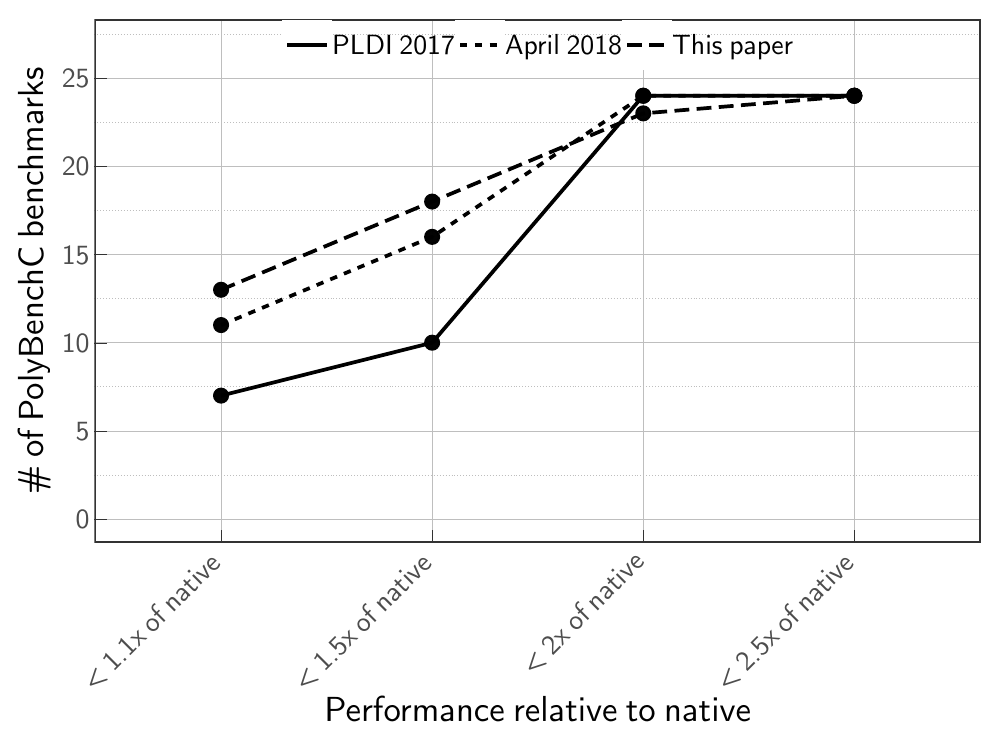}
  \caption{Number of PolyBenchC benchmarks performing within
  $x$$\times$ of native. In 2017~\cite{haas:2017:webassembly},
  seven benchmarks performed within 1.1$\times$ of native.
  In April 2018, we found that 11 performed within 1.1$\times$ of native. In
  May 2019, 13 performed with 1.1$\times$ of native.}
  \label{fig:polybench-improvements}
\end{figure}

A major goal of \wasm{} is to be faster than JavaScript. For example, the paper
that introduced \wasm{}~\cite{haas:2017:webassembly} showed that when a C
program is compiled to \wasm{} instead of JavaScript (\asmjs), it runs 34\%
faster in Google Chrome. That paper also showed that the performance of \wasm{}
is competitive with native code: of the 24 benchmarks evaluated, the running
time of seven benchmarks using \wasm{} is within 10\% of native code, and
almost all of them are less than $2\times$ slower than native code.
Figure~\ref{fig:polybench-improvements} shows that \wasm{} implementations have
continuously improved with respect to these benchmarks.
In 2017, only seven benchmarks performed within 1.1$\times$ of native,
but by 2019, this number increased to 13.

These results appear promising, but they beg the question: \emph{are these
24 benchmarks representative of \wasm{}'s intended use cases}?

\paragraph{The Challenge of Benchmarking \wasm{}}

The aforementioned suite of 24 benchmarks is the PolybenchC benchmark
suite~\cite{polybench}, which is designed to measure the effect of
polyhedral loop optimizations in compilers. All the benchmarks in the
suite are small scientific computing kernels rather than full
applications (e.g., matrix multiplication and LU Decomposition); each is
roughly 100 LOC. While \wasm{} is designed to accelerate scientific
kernels on the Web, it is also explicitly designed for a much richer set
of full applications.

The \wasm{} documentation highlights several intended use
cases~\cite{wasm-use-cases}, including scientific kernels, image editing,
video editing, image recognition, scientific visualization, simulations,
programming language interpreters, virtual machines, and POSIX applications.
Therefore, \wasm{}'s strong performance on the scientific kernels in PolybenchC
do not imply that it will perform well given a different kind of application.

We argue that a more comprehensive evaluation of \wasm{} should rely on an
established benchmark suite of large programs, such as the SPEC CPU benchmark
suites. In fact, the SPEC CPU 2006 and 2017 suite of
benchmarks include several applications that fall under the intended use cases of
\wasm{}: eight benchmarks are scientific applications (e.g., \texttt{433.milc},
\texttt{444.namd}, \texttt{447.dealII}, \texttt{450.soplex}, and
\texttt{470.lbm}), two benchmarks involve image and video processing
(\texttt{464.h264ref} and \texttt{453.povray}), and all of the benchmarks are POSIX
applications.

Unfortunately, it is not possible to simply compile a sophisticated
native program to \wasm{}. Native programs, including the programs in
the SPEC CPU suites, require operating system services, such as a
filesystem, synchronous I/O, and processes, which \wasm{} and the
browser do not provide. The SPEC benchmarking harness itself requires
a file system, a shell, the ability to spawn processes, and other Unix
facilities. To overcome these limitations when porting native
applications to the web, many programmers painstakingly modify their
programs to avoid or mimic missing operating system
services. Modifying well-known benchmarks, such as SPEC CPU, would not
only be time consuming but would also pose a serious threat to
validity.

The standard approach to running these applications today is to use
Emscripten, a toolchain for compiling C and C++ to
WebAssembly~\cite{emscripten}. Unfortunately, Emscripten only supports
the most trivial system calls and does not scale up to large-scale
applications. For example, to enable applications to use synchronous
I/O, the default Emscripten \texttt{MEMFS} filesystem loads the entire
filesystem image into memory before the program begins executing. For
SPEC, these files are too large to fit into memory.

A promising alternative is to use \browsix, a framework that enables
running unmodified, full-featured Unix applications in the
browser~\cite{browsix-web,powers:2017:browsix}. \Browsix implements
a Unix-compatible kernel in JavaScript, with full support for
processes, files, pipes, blocking I/O, and other Unix features.
Moreover, it includes a C/C++ compiler (based on Emscripten)
that allows programs to run in the browser
unmodified. The \Browsix case studies include complex applications,
such as \LaTeX, which runs entirely in the browser without any
source code modifications.

Unfortunately, \browsix is a JavaScript-only solution, since it was
 built before the release of
\wasm{}. Moreover, \browsix suffers from high performance overhead,
which would be a significant confounder while benchmarking. Using \browsix,
it would be difficult to tease apart the poorly performing benchmarks
from performance degradation introduced by \browsix.

\subsection*{Contributions}

\begin{itemize}
\item \textbf{\BrowsixWasm:} We develop \BrowsixWasm, a significant
  extension to and enhancement of \Browsix that allows us to
  compile Unix programs to \wasm{} and run them in the browser with
  no modifications. In
  addition to integrating functional extensions, \BrowsixWasm
  incorporates performance optimizations that drastically improve its
  performance, ensuring that CPU-intensive applications operate with
  virtually no overhead imposed by \BrowsixWasm ($\S\ref{sec:browsix_wasm}$).

\item \textbf{\BrowsixSPEC:} We develop \BrowsixSPEC, a harness
  that extends \BrowsixWasm to allow automated collection of detailed
  timing and hardware on-chip performance counter information in order
  to perform detailed measurements of application performance ($\S\ref{sec:system:bench-setup}$).

\item \textbf{Performance Analysis of WebAssembly:} Using
  \BrowsixWasm and \BrowsixSPEC, we conduct the first comprehensive
  performance analysis of \wasm{} using the SPEC
  CPU benchmark suite (both 2006 and 2017). This evaluation confirms
  that \wasm{} does run faster than JavaScript (on average 1.3$\times$
  faster across SPEC CPU). However, contrary to prior work, we find
  a substantial gap between
  \wasm{} and native performance: code compiled to \wasm{}
  runs on average 1.55$\times$ slower
  in Chrome and
  1.45$\times$ slower in Firefox than native code ($\S\ref{sec:eval}$).

\item \textbf{Root Cause Analysis and Advice for Implementers:} We
  conduct a forensic analysis with the aid of performance counter
  results to identify the root causes of this performance gap. We find
  the following results: 
  \begin{enumerate} 
    \item The instructions produced by \wasm{} have more
  loads and stores than native code (2.02$\times$ more loads and
  2.30$\times$ more stores in Chrome; 1.92$\times$ more loads and
  2.16$\times$ more stores in Firefox).  We attribute this to reduced
  availability of registers, a sub-optimal register allocator, and a
  failure to effectively exploit a wider range of x86 addressing
  modes.
  \item The instructions produced by \wasm{}
  have more branches, because \wasm{} requires several dynamic safety checks.
  \item Since \wasm{} generates more instructions, it leads to more L1 instruction cache misses.
  \end{enumerate}
  We provide guidance to help WebAssembly
  implementers focus their optimization efforts in order to close
  the performance gap between WebAssembly and native code ($\S\ref{sec:ex},\ref{sec:spec}$).

\end{itemize}

\browsixWasm and \BrowsixSPEC are available at
\url{https://browsix.org}.

\begin{figure*}
\centering
\includegraphics[width=0.9\textwidth]{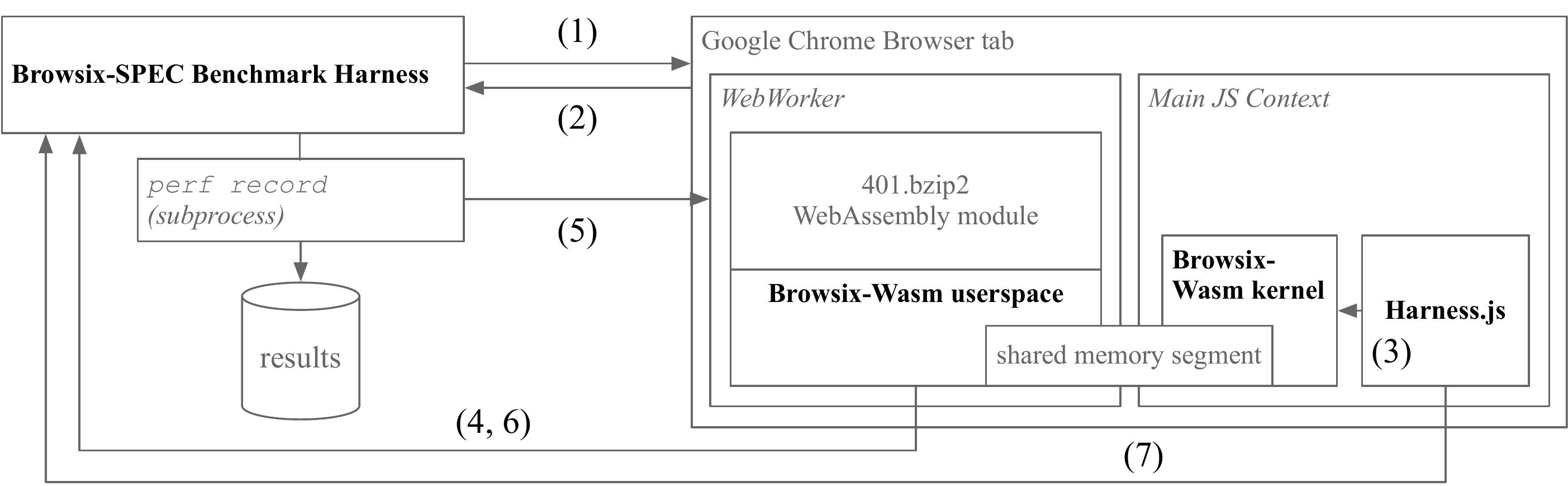}
\caption{The framework for running SPEC benchmarks in browsers.
    \textbf{Bold} components are new or heavily modified
    (\S\ref{sec:system:bench-setup}).}
\label{fig:architecture}
\end{figure*}

\section{From \Browsix to \browsixWasm}
\label{sec:browsix_wasm}

\Browsix~\cite{powers:2017:browsix} mimics a Unix kernel within the browser and
includes a compiler (based on Emscripten~\cite{emscripten,asmjs:announcement})
that compiles native programs to JavaScript. Together, they allow native
programs (in C, C++, and Go) to run in the browser and freely use operating
system services, such as pipes, processes, and a filesystem. However, \Browsix
has two major limitations that we must overcome. First, \Browsix compiles
native code to JavaScript and not \wasm. Second, the \Browsix kernel has
significant performance issues. In particular, several common system calls have
very high overhead in \Browsix, which makes it hard to compare the performance
of a program running in \Browsix to that of a program running natively. We address
these limitations by building a new in-browser kernel called \browsixWasm,
which supports \wasm programs and eliminates the performance bottlenecks of
\browsix.

\paragraph{Emscripten Runtime Modifications}

\Browsix modifies the Emscripten compiler to allow processes (which run in
WebWorkers) to communicate with the \Browsix kernel (which runs on the main
thread of a page). Since \Browsix compiles native programs to JavaScript, this
is relatively straightforward: each process' memory is a buffer that is shared
with the kernel (a \texttt{SharedArrayBuffer}), thus system calls can directly
read and write process memory. However, this approach has two significant
drawbacks. First, it precludes growing the heap on-demand; the shared memory
must be sized large enough to meet the high-water-mark heap size of the
application for the entire life of the process. Second, JavaScript contexts
(like the main context and each web worker context) have a fixed limit on their
heap sizes, which is currently approximately 2.2 GB in Google
Chrome~\cite{js-heap-limit}. This cap imposes a serious limitation on running
multiple processes: if each process reserves a 500 MB heap, \Browsix would only
be able to run at most four concurrent processes.
A deeper problem is that \wasm memory cannot be shared across WebWorkers and
does not support the \texttt{Atomic} API, which \browsix processes use to wait
for system calls.

\BrowsixWasm uses a different approach to process-kernel communication that is
also faster than the \Browsix approach. \BrowsixWasm modifies the Emscripten runtime
system to create an auxiliary buffer (of 64MB) for each process that is shared with the
kernel, but is distinct from process memory. Since this auxiliary buffer 
is a \texttt{SharedArrayBuffer} the \BrowsixWasm process and kernel 
can use \texttt{Atomic} API for communication.
When a system call references
strings or buffers in the process's heap (e.g., \texttt{writev} or
\texttt{stat}), its runtime system copies data from the process memory to the
shared buffer and sends a message to the kernel with locations of the copied 
data in auxiliary memory.
Similarly, when a system call writes data to the auxiliary 
buffer (e.g., \texttt{read}), its runtime system copies the data from 
the shared buffer to the process memory at the memory specified. 
Moreover, if a system call specifies a buffer in process memory 
for the kernel to write to (e.g., \texttt{read}), the runtime allocates a 
corresponding buffer in auxiliary memory and passes it to the kernel. 
In case the system call is either reading 
or writing data of size more than 64MB, \Browsixwasm divides this call into
several calls such that each call only reads or writes at maximum 64MB
of data. The cost of these memory copy operations is dwarfed by the overall cost of
the system call invocation, which involves sending a message between process
and kernel JavaScript contexts. We show in \S\ref{sec:browsix-overhead} that
\BrowsixWasm has negligible overhead.

\paragraph{Performance Optimization}
While building \BrowsixWasm and doing our preliminary performance evaluation,
we discovered several performance issues in parts of the \Browsix kernel.
Left unresolved, these performance issues would be a threat to the validity
of a performance comparison between \wasm and native code.
The most serious case was in the shared filesystem
component included with \Browsix/\BrowsixWasm, \BrowserFS. Originally,
on each append operation on a file, \BrowserFS would allocate a new,
larger buffer, copying the previous and new contents into the new
buffer. Small appends could impose substantial performance
degradation. Now, whenever a buffer backing a file requires additional
space, \BrowserFS grows the buffer by at least 4 KB.  This change
alone decreased the time the \texttt{464.h264ref} benchmark spent in
\Browsix from 25 seconds to under 1.5 seconds. We made a series of
improvements that reduce overhead throughout \BrowsixWasm. Similar, if
less dramatic, improvements were made to reduce the number of
allocations and the amount of copying in the kernel implementation of
pipes.

\section{\BrowsixSPEC}\label{sec:system:bench-setup}

To reliably execute WebAssembly benchmarks while capturing performance counter data, we developed \BrowsixSpec.
\BrowsixSpec works with \BrowsixWasm to manage spawning browser instances, serving benchmark assets (e.g., the compiled WebAssembly programs and test inputs), spawning \texttt{perf} processes to record performance counter data, and validating benchmark outputs.

We use \BrowsixSpec to run three benchmark suites to evaluate WebAssembly's performance: SPEC CPU2006, SPEC CPU2017, and PolyBenchC.
These benchmarks are compiled to native code using Clang 4.0, and WebAssembly using \BrowsixWasm.
We made no modifications to Chrome or Firefox, and the browsers are run with their standard sandboxing and isolation features enabled.
\BrowsixWasm is built on top of standard web platform features and requires no direct access to host resources -- instead, benchmarks make standard HTTP requests to \BrowsixSpec.

\subsection{\BrowsixSpec Benchmark Execution}

Figure~\ref{fig:architecture} illustrates the key pieces of \BrowsixSpec in play when running a benchmark, such as \texttt{401.bzip2} in Chrome.
First (1), the \BrowsixSpec benchmark harness launches a new browser instance using a WebBrowser automation tool, Selenium.\footnote{\url{https://www.seleniumhq.org/}}
(2) The browser loads the page's HTML, harness JS, and \BrowsixWasm kernel JS over HTTP from the benchmark harness.
(3) The harness JS initializes the \BrowsixWasm kernel and starts a new \BrowsixWasm process executing the \texttt{runspec} shell script (not shown in Figure \ref{fig:architecture}).
\texttt{runspec} in turn spawns the standard \texttt{specinvoke} (not shown), compiled from the C sources provided in SPEC 2006.
\texttt{specinvoke} reads the \texttt{speccmds.cmd} file from the \BrowsixWasm filesystem and starts \texttt{401.bzip2} with the appropriate arguments.
(4) After the WebAssembly module has been instantiated but before the benchmark's \texttt{main} function is invoked, the \browsixwasm userspace runtime does an XHR request to \BrowsixSpec to begin recording performance counter stats.
(5) The benchmark harness finds the Chrome thread corresponding to the Web Worker \texttt{401.bzip2} process and attaches \texttt{perf} to the process.
(6) At the end of the benchmark, the \browsixwasm userspace runtime does a final XHR to the benchmark harness to end the \texttt{perf record} process.
When the \texttt{runspec} program exits (after potentially invoking the test binary several times), the harness JS POSTs (7) a \texttt{tar} archive of the SPEC results directory to \BrowsixSpec.
After \BrowsixSpec receives the full results archive, it unpacks the results to a temporary directory and validates the output using the \texttt{cmp} tool provided with SPEC 2006.
Finally, \BrowsixSpec kills the browser process and records the benchmark results.

%
%
%
%
%

\begin{figure*}
  \begin{subfigure}[b]{0.5\textwidth}
    \includegraphics[width=\linewidth]{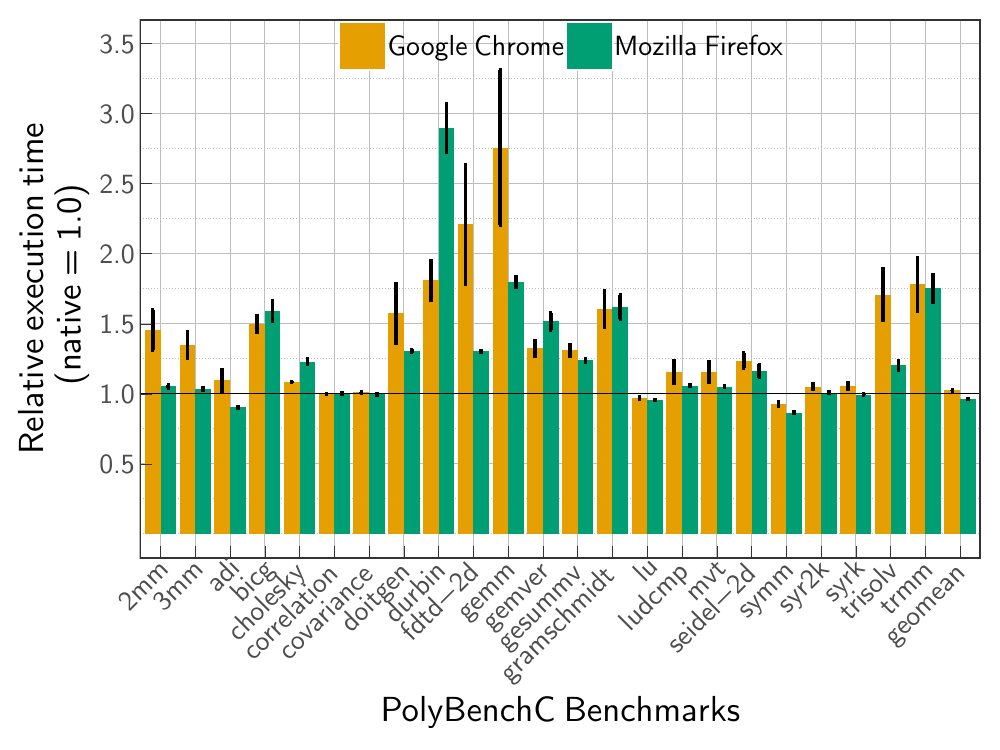}
    \caption{
    \label{fig:polybench-results}
    }
    
  \end{subfigure}
  \begin{subfigure}[b]{0.5\textwidth}
    \includegraphics[width=\linewidth]{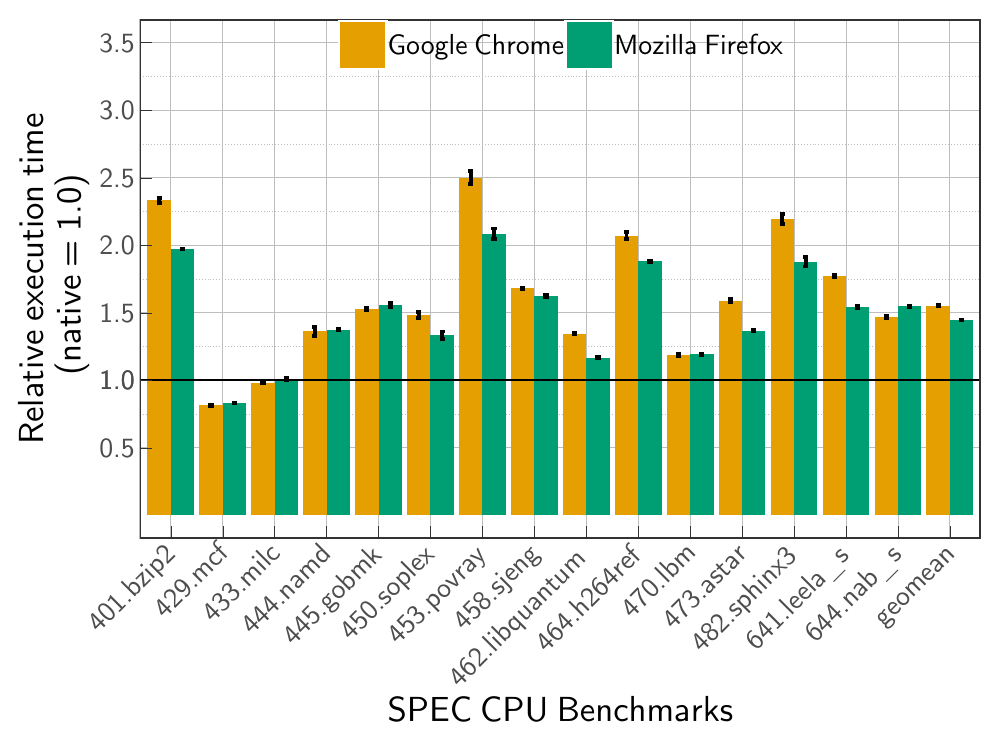}
  \caption{
    \label{fig:spec-relative-times}
    }
    
  \end{subfigure}  
  \caption{The performance of the PolyBenchC and the SPEC CPU
    benchmarks compiled to \wasm{} (executed in Chrome and Firefox)
    relative to native, using \BrowsixWasm and \BrowsixSPEC. The SPEC
    CPU benchmarks exhibit higher overhead overall than the PolyBenchC
    suite, indicating a significant performance gap exists between
    WebAssembly and native.}
    \vspace{-1em}
\end{figure*}
\section{Evaluation}
\label{sec:eval}

We use \BrowsixWasm{} and \BrowsixSPEC{} to evaluate the performance of \wasm{}
using three benchmark suites: SPEC CPU2006, SPEC CPU2017, and PolyBenchC. We
include PolybenchC benchmarks for comparison with the original \wasm{}
paper~\cite{haas:2017:webassembly}, but argue that these benchmarks do not represent typical
workloads. The SPEC benchmarks are representative and require \BrowsixWasm{} to
run successfully. We run all benchmarks on a 6-Core Intel Xeon E5-1650 v3 CPU with
hyperthreading and 64 GB of RAM running Ubuntu 16.04 with Linux kernel v4.4.0.
We run all benchmarks using two state-of-the-art browsers: Google Chrome 74.0
and Mozilla Firefox 66.0. We compile benchmarks to native code using Clang
4.0\footnote{The flags to Clang are \texttt{-O2 -fno-strict-aliasing}.} and
to \wasm{} using \BrowsixWasm (which is based on Emscripten with Clang 4.0).\footnote{\BrowsixWasm runs Emscripten with the flags \texttt{-O2 -s
TOTAL\_MEMORY=1073741824 -s ALLOW\_MEMORY\_GROWTH=1 -fno-strict-aliasing}.}
Each benchmark was executed five times. We report the average of all running 
times and the standard error. The execution time measured is the difference 
between wall clock time when the program starts, i.e. after WebAssembly JIT compilation 
concludes, and when the program ends.

\subsection{PolyBenchC Benchmarks}

Haas et al.~\cite{haas:2017:webassembly} used PolybenchC to benchmark \wasm{}
implementations because the PolybenchC benchmarks do not make system calls.
As we have already argued, the PolybenchC benchmarks are small
scientific kernels that are typically used to benchmark polyhedral optimization
techniques, and do not represent larger applications. Nevertheless, it is still
valuable for us to run PolybenchC with \BrowsixWasm{}, because it demonstrates
that our infrastructure for system calls does not have any overhead.
 Figure~\ref{fig:polybench-results} shows the execution
time of the PolyBenchC benchmarks in  \BrowsixWasm{} and when run natively.
We are able to reproduce the majority of the results from the original
WebAssembly paper~\cite{haas:2017:webassembly}. We find that
\BrowsixWasm imposes a very low overhead: an average of 0.2\% and a maximum of 1.2\%.

\subsection{SPEC Benchmarks}

We now evaluate \BrowsixWasm{} using the C/C++ benchmarks from SPEC CPU2006 and
SPEC CPU2017 (the new C/C++ benchmarks and the speed benchmarks), which use
system calls extensively. We exclude four data points that either do not
compile to \wasm{}\footnote{\texttt{400.perlbench}, \texttt{403.gcc},
\texttt{471.omnetpp}, and \texttt{456.hmmer} from
SPEC CPU2006 do not compile with Emscripten.} or allocate more memory than
\wasm{} allows.\footnote{From SPEC CPU2017, the \texttt{ref} dataset of
\texttt{638.imagick\_s} and \texttt{657.xz\_s} require more than 4 GB RAM.
However, these benchmarks do work with their \texttt{test} dataset.}
Table~\ref{tab:spec-absolute-times}
shows the absolute execution times of the SPEC benchmarks when running
with \BrowsixWasm in both Chrome and Firefox, and when running natively.

\wasm{} performs worse than native for all benchmarks except for \texttt{429.mcf}
and \texttt{433.milc}.
In Chrome, \wasm{}'s maximum overhead is
 2.5$\times$ over native and 7 out of 15 benchmarks
have a running time within  1.5$\times$ of native. 
In Firefox, \wasm{} is within 2.08$\times$ of native and performs within 1.5$\times$ of native for 7 out of 15 benchmarks.
On average, \wasm{}
is 1.55$\times$ slower than
native in Chrome, and 1.45$\times$ slower than native in Firefox.
Table~\ref{tab:spec-compile-times} shows the time required to compile the SPEC
benchmarks using Clang and Chrome. (To the best of our knowledge, Firefox cannot
report \wasm{} compile times.) In all cases, the compilation time is negligible compared
to the execution time. However, the Clang compiler is orders of magnitude slower
than the \wasm{} compiler. Finally, note that Clang compiles benchmarks from C++
source code, whereas Chrome compiles \wasm{}, which is a simpler format than C++.


  
\begin{table}[t]
  \small
\begin{tabular}{l|r|r|r}
\textbf{Benchmark} & \textbf{Native} & \textbf{\thead{Google\\ Chrome}}& \textbf{\thead{Mozilla\\Firefox}}\\ \hline
\texttt{401.bzip2} & 370 $\pm$ 0.6 & 864 $\pm$ 6.4 & 730 $\pm$ 1.3\\
\texttt{429.mcf} & 221 $\pm$ 0.1 & 180 $\pm$ 0.9 & 184 $\pm$ 0.6\\
\texttt{433.milc} & 375 $\pm$ 2.6 & 369 $\pm$ 0.5 & 378 $\pm$ 0.6\\
\texttt{444.namd} & 271 $\pm$ 0.8 & 369 $\pm$ 9.1 & 373 $\pm$ 1.8\\
\texttt{445.gobmk} & 352 $\pm$ 2.1 & 537 $\pm$ 0.8 & 549 $\pm$ 3.3\\
\texttt{450.soplex} & 179 $\pm$ 3.7 & 265 $\pm$ 1.2 & 238 $\pm$ 0.5\\
\texttt{453.povray} & 110 $\pm$ 1.9 & 275 $\pm$ 1.3 & 229 $\pm$ 1.5\\
\texttt{458.sjeng} & 358 $\pm$ 1.4 & 602 $\pm$ 2.5 & 580 $\pm$ 2.0\\
\texttt{462.libquantum} & 330 $\pm$ 0.8 & 444 $\pm$ 0.2 & 385 $\pm$ 0.8\\
\texttt{464.h264ref} & 389 $\pm$ 0.7 & 807 $\pm$ 11.0 & 733 $\pm$ 2.4\\
\texttt{470.lbm} & 209 $\pm$ 1.1 & 248 $\pm$ 0.3 & 249 $\pm$ 0.5\\
\texttt{473.astar} & 299 $\pm$ 0.5 & 474 $\pm$ 3.5 & 408 $\pm$ 1.0\\
\texttt{482.sphinx3} & 381 $\pm$ 7.1 & 834 $\pm$ 1.8 & 713 $\pm$ 3.6\\
\texttt{641.leela\_s} & 466 $\pm$ 2.7 & 825 $\pm$ 4.6 & 717 $\pm$ 1.2\\
\texttt{644.nab\_s} & 2476 $\pm$ 11 & 3639 $\pm$ 5.6 & 3829 $\pm$ 6.7\\ \hline
\textbf{Slowdown: geomean} & -- & \textbf{1.55}$\times$ & \textbf{1.45}$\times$\\
\textbf{Slowdown: median} & -- & \textbf{1.53}$\times$ & \textbf{1.54}$\times$\\
\end{tabular}
\caption{Detailed breakdown of SPEC CPU benchmarks execution times
  (of 5 runs) for native (Clang) and \wasm{} (Chrome and Firefox); all times are in seconds. The median slowdown of \wasm is
  $1.53\times$ for Chrome and $1.54\times$ for Firefox.
\label{tab:spec-absolute-times}}
\vspace{-1em}
\end{table}

\begin{table}[t]
\begin{tabular}{l|r|r|r}
\textbf{Benchmark} & \textbf{Clang 4.0} & \textbf{\thead{Google Chrome}}\\ \hline
\texttt{401.bzip2}&1.9 $\pm$ 0.018 &0.53 $\pm$ 0.005\\
\texttt{429.mcf}&0.3 $\pm$ 0.003&0.15 $\pm$ 0.005\\
\texttt{433.milc}&2.2 $\pm$ 0.02&0.3 $\pm$ 0.003\\
\texttt{444.namd}&4.6 $\pm$ 0.02&0.78 $\pm$ 0.004\\
\texttt{445.gobmk}&12.1 $\pm$ 0.2&1.4 $\pm$ 0.014\\
\texttt{450.soplex}&6.9 $\pm$ 0.01&1.2 $\pm$ 0.009\\
\texttt{453.povray}&15.3 $\pm$ 0.03&1.2 $\pm$ 0.012\\
\texttt{458.sjeng}&1.9 $\pm$ 0.01&0.35 $\pm$ 0.001\\
\texttt{462.libquantum}&6.9 $\pm$ 0.03&0.15 $\pm$ 0.002\\
\texttt{464.h264ref}&10.3 $\pm$ 0.06&1.0 $\pm$ 0.03\\
\texttt{470.lbm}&0.3 $\pm$ 0.001&0.14 $\pm$ 0.004\\
\texttt{473.astar}&0.73 $\pm$ 0.005&0.24 $\pm$ 0.004\\
\texttt{482.sphinx3}&3.0 $\pm$ 0.04&0.48 $\pm$ 0.007\\
\texttt{641.leela\_s}&4.3 $\pm$ 0.05&0.74 $\pm$0.003\\
\texttt{644.nab\_s}&4.1 $\pm$ 0.03&0.41 $\pm$0.001\\
\end{tabular}
\caption{Compilation times of SPEC CPU benchmarks
  (average of 5 runs) for Clang 4.0 and \wasm{} (Chrome); all times are in seconds.
\label{tab:spec-compile-times}}
\vspace{-1em}
\end{table}

\subsubsection{\BrowsixWasm Overhead}
\label{sec:browsix-overhead}

\begin{figure}[!t]
  \includegraphics[width=\linewidth]{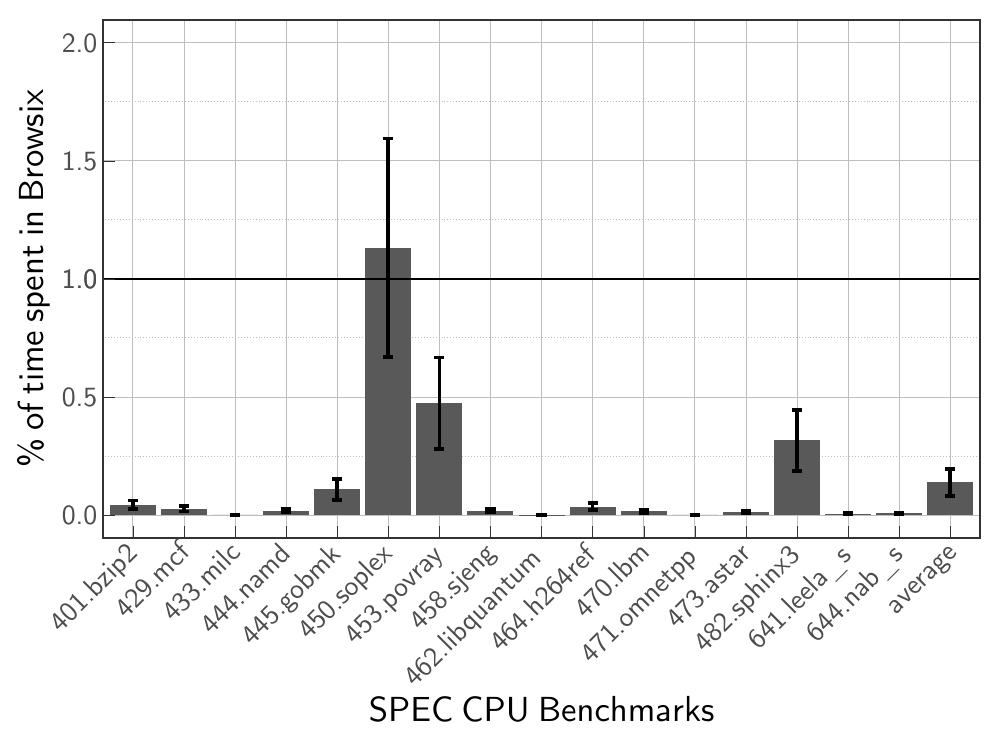}
  \vspace{-1em}
  \caption{Time spent (in \%) in \BrowsixWasm calls in Firefox for 
    SPEC benchmarks compiled to \wasm{}. \BrowsixWasm imposes a
    mean overhead of only 0.2\%.}
  \label{fig:browsix-overhead}
  \vspace{-1em}
\end{figure}

It is important to rule out the possibility that the slowdown that we report is
due to poor performance in our implementation of \BrowsixWasm{}. In particular,
\BrowsixWasm{} implements system calls without modifying the browser, and
system calls involve copying data (\S\ref{sec:browsix_wasm}), which may be
costly. To quantify the overhead of \BrowsixWasm, we instrumented its system
calls to measure all time spent in \BrowsixWasm.
Figure~\ref{fig:browsix-overhead} shows the percentage of time spent in
\BrowsixWasm in Firefox using the SPEC benchmarks. For 14 of the 15
benchmarks, the overhead is less than 0.5\%. The maximum overhead is 1.2\%.
On average, the overhead of \BrowsixWasm is only 0.2\%. Therefore, we conclude that \BrowsixWasm{} has negligible overhead and
does not substantially affect the performance counter results of programs executed in
\wasm{}.

\subsubsection {Comparison of \wasm{} and \asmjs{}}

\begin{figure}[!t]
  \includegraphics[width=\linewidth]{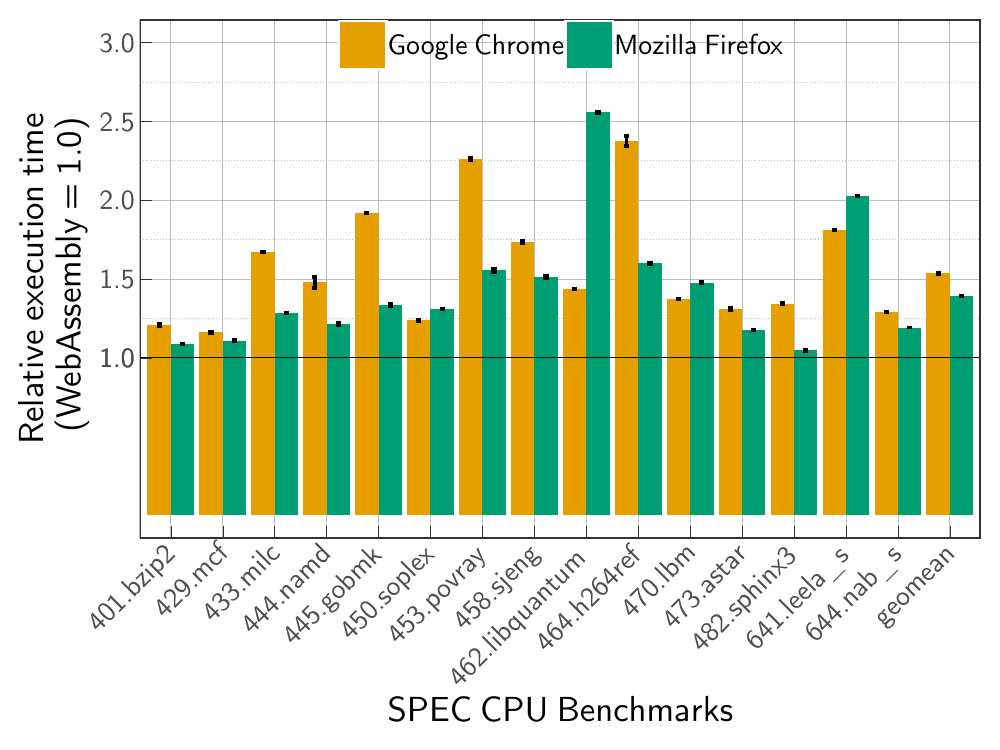}
  \caption{Relative time of \asmjs{} to 
           \wasm{} for Chrome and Firefox. \wasm{} is 1.54$\times$ faster than \asmjs{} in Chrome and 1.39$\times$ faster
           than \asmjs{} in Firefox.}
  \label{fig:asmjs-wasm-relative}
  \vspace{-1em}
\end{figure}

A key claim in the original work on \wasm{} was that it is significantly
faster than \asmjs. We now test that claim using the SPEC
benchmarks. For this comparison, we modified \browsixWasm{} to also support
processes compiled to \asmjs. The alternative would have been to benchmark
the \asmjs processes using the original \browsix. However, as we discussed
earlier, \browsix has performance problems that would have been a threat to
the validity of our results.
Figure~\ref{fig:asmjs-wasm-relative} shows the speedup of the SPEC benchmarks
using \wasm, relative to their running time using \asmjs using both
Chrome and Firefox. \wasm{} outperforms \asmjs{} in both browsers: the mean
speedup is 1.54$\times$ in Chrome and 1.39$\times$ in Firefox.

\begin{figure}[!t]
  \includegraphics[width=\linewidth]{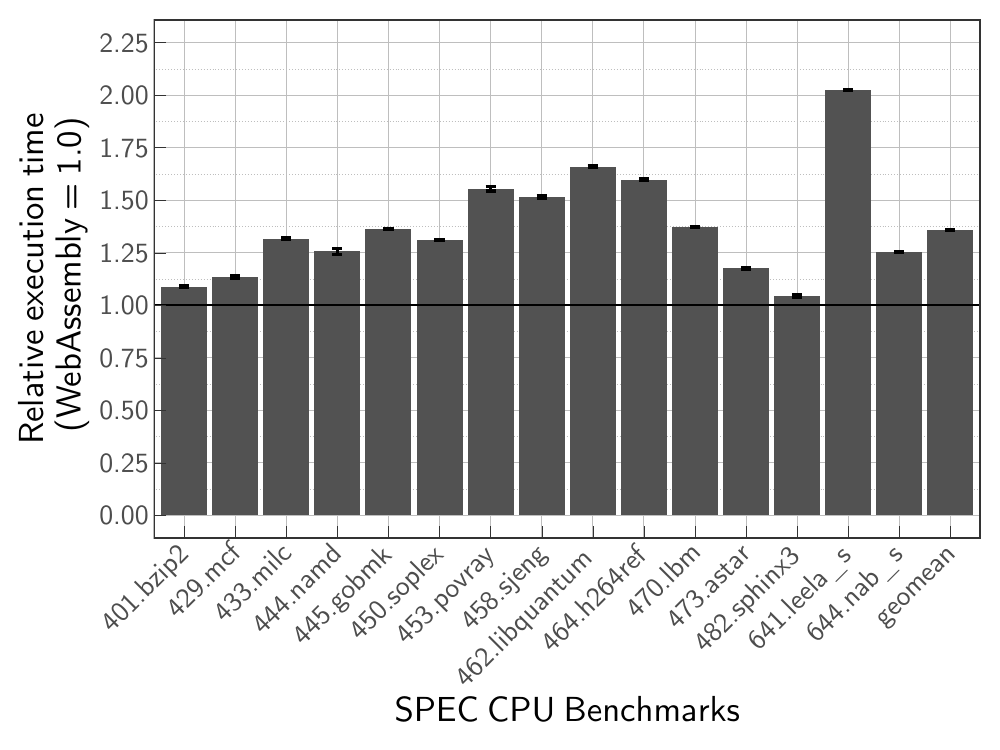}
  \caption{Relative \emph{best} time of \asmjs{} to the best 
           time of \wasm{}. \wasm{} is 1.3$\times$ faster than \asmjs{}.}
  \label{fig:asmjs-wasm-best-relative}
  \vspace{-1em}
\end{figure}

Since the performance difference between Chrome  and Firefox is substantial, in
Figure~\ref{fig:asmjs-wasm-best-relative} we show the speedup of each
benchmark by selecting the best-performing browser for \wasm{} and the
best-performing browser of \asmjs{} (i.e., they may be different browsers).
These results show that \wasm{} consistently performs better than \asmjs,
with a mean speedup of 1.3$\times$. Haas et al.~\cite{haas:2017:webassembly}
also found that \wasm{} gives a mean speedup of 1.3$\times$ over \asmjs using
PolyBenchC.

\section {Case Study: Matrix Multiplication}
\label{sec:ex}
In this section, we illustrate the performance differences between
\wasm{} and native code using a C function that performs matrix
multiplication, as shown in Figure~\ref{fig:mat-mult}.  Three matrices
are provided as arguments to the function, and the results of
\texttt{A} ($N_I\times N_K$) and \texttt{B} ($N_K \times N_J$) are
stored in \texttt{C} ($N_I \times N_J$), where $N_I, N_K, N_J$ are
constants defined in the program.

\begin{figure}[h]
  \includegraphics[width=0.9\columnwidth]{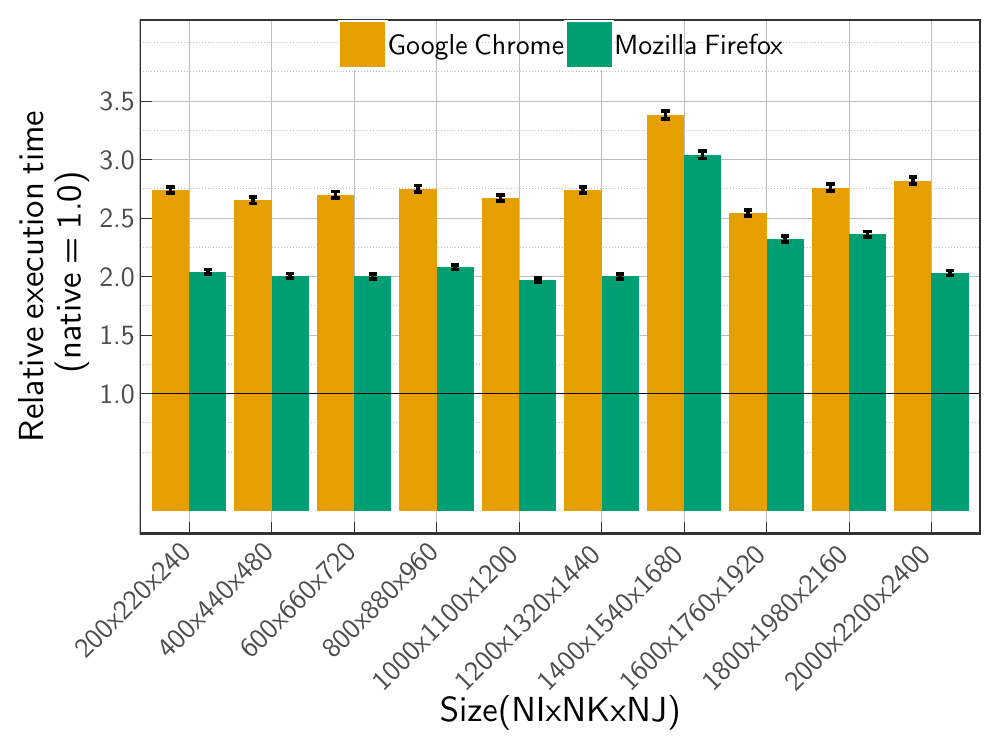}
  \vspace{-0.5em}
  \caption{Performance of \wasm in Chrome and Firefox for
  different matrix sizes relative to native code. \wasm{} is always 
  between 2$\times$ to 3.4$\times$ slower than native. \label{fig:matmul-perf}}
  \vspace{-2em}
\end{figure}

In \wasm, this function is $2\times$--$3.4\times$ slower than native in both
Chrome and Firefox with a variety of matrix sizes
(Figure~\ref{fig:matmul-perf}). We compiled the function with \texttt{-O2} and
disabled automatic vectorization, since \wasm does not support vectorized
instructions.

\begin{figure*}[!t]
\begin{minipage}{\columnwidth}
\begin{subfigure}[t]{\linewidth}
\vspace{-1em}
\begin{lstlisting}[language=myC]
void matmul (int C[NI][NJ],
             int A[NI][NK], 
             int B[NK][NJ]) {
  for (int i = 0; i < NI; i++) {
    for (int k = 0; k < NK; k++) {
      for (int j = 0; k < NJ; j++) {
        C[i][j] += A[i][k] * B[k][j]; 
      }
    }
  }
}
\end{lstlisting}
\vspace{-1em}
\caption{\texttt{matmul} source code in C.}
\label{fig:mat-mult}
\vspace{1em}
\end{subfigure}

\begin{subfigure}[t]{\linewidth} 
\begin{lstlisting}[language=x8664intel, escapechar=|]
xor  r8d, r8d           #i <- 0
L1:                     #start first loop |\label{line:clang-code:l1-start}|
  mov  r10, rdx
  xor  r9d, r9d         #k <- 0
  L2:                   #start second loop |\label{line:clang-code:l2-start}|
    imul  rax, 4*NK, r8
    add   rax, rsi
    lea   r11, [rax + r9*4]
    mov   rcx, -NJ     #j <- -NJ |\label{line:clang-code:movq-rcx}|
    L3:                 #start third loop |\label{line:clang-code:l3-start}|
      mov   eax, [r11]
      mov   ebx, [r10 + rcx*4 + 4400]
      imul  ebx, eax    
      add   [rdi + rcx*4 + 4*NJ], ebx |\label{line:clang-code:addl}|
      add   rcx, 1      #j <- j + 1 |\label{line:clang-code:rcx+1}|
    jne L3              #end third loop       |\label{line:clang-code:jmp-l3}|

    add   r9,  1        #k <- k + 1
    add   r10, 4*NK
    cmp   r9,  NK     
  jne L2                #end second loop |\label{line:clang-code:l2-end}|

  add  r8,  1           #i <- i + 1
  add  rdi, 4*NJ
  cmp  r8,  NI
jne L1                  #end first loop |\label{line:clang-code:l1-end}|
pop  rbx
ret
\end{lstlisting}
\caption{Native x86-64 code for \texttt{matmul} generated by Clang.\label{fig:clang-code}}
\end{subfigure}
\end{minipage}
\quad\quad
\begin{minipage}{0.95\columnwidth}
\begin{subfigure}[t]{\linewidth}
\begin{lstlisting}[language=x8664intel, xleftmargin=0.1cm, escapechar=|, numbers=left]
mov [rbp-0x28],rax                    |\label{line:v8-code:rax-init}|
mov [rbp-0x20],rdx                    |\label{line:v8-code:rdx-init}|
mov [rbp-0x18],rcx                    |\label{line:v8-code:rcx-init}|
xor edi,edi               #i <- 0 
jmp L1'                               |\label{line:v8-code:jmp-L1'}|
L1:                       #start first loop
  mov ecx,[rbp-0x18]                |\label{line:v8-code:l1-start}| |\label{line:v8-code:spill1}|
  mov edx,[rbp-0x20]                |\label{line:v8-code:spill2}|
  mov eax,[rbp-0x28]                |\label{line:v8-code:spill3}|
  L1': 
  imul r8d,edi,0x1130
  add r8d,eax
  imul r9d,edi,0x12c0
  add r9d,edx
  xor r11d,r11d           #k <- 0
  jmp L2'                           |\label{line:v8-code:jmp-L2'}|
  L2:                     #start second loop
    mov ecx,[rbp-0x18]            |\label{line:v8-code:l2-start}|
    L2':
    imul r12d,r11d,0x1130         
    lea r14d,[r9+r11*4]
    add r12d,ecx
    xor esi,esi           #j <- 0
    mov r15d,esi
    jmp L3'                       |\label{line:v8-code:jmp-L3'}|
    L3:                   #start third loop 
      mov r15d,eax           |\label{line:v8-code:l3-start}|
      L3': 
      lea eax,[r15+0x1]   #j <- j + 1 |\label{line:v8-code:rax+1}|
      lea edx,[r8+r15*4]
      lea r15d,[r12+r15*4]
      mov esi,[rbx+r14*1]
      mov r15d,[rbx+r15*1]
      imul r15d,esi
      mov ecx,[rbx+rdx*1]   |\label{line:v8-code:addl-start}|
      add ecx,r15d             
      mov [rbx+rdx*1],ecx    |\label{line:v8-code:addl-end}|    
      cmp eax,NJ          #j < NJ |\label{line:v8-code:cmp-rax}|
    jnz L3                #end third loop |\label{line:v8-code:jmp-l3}|
  add r11,0x1             #k++
  cmp r11d,NK             #k < NK
  jnz L2                  #end second loop |\label{line:v8-code:l2-end}|     
add edi,0x1               #i++
cmp edi,NI                #i < NI    
jnz L1                    #end first loop |\label{line:v8-code:l1-end}| 
retl
\end{lstlisting}
\vspace{-1em}
\caption{x86-64 code JITed by Chrome from \wasm \texttt{matmul}.\label{fig:v8-code}}
\end{subfigure}
\end{minipage}
\caption{Native code for \texttt{matmul} is shorter, has less register pressure, and fewer branches than the code JITed by Chrome.
  \S\ref{sec:spec} shows that these inefficiencies are pervasive, reducing performance across the SPEC CPU benchmark suites.}
\end{figure*}

Figure~\ref{fig:clang-code} shows native code generated for the \texttt{matmul} 
function by \texttt{clang-4.0}.
Arguments are passed to the function in the \texttt{rdi}, \texttt{rsi}, and 
\texttt{rdx} registers, as specified in the System V AMD64 ABI calling 
convention~\cite{linux-calling-conventions}.
Lines~\ref{line:clang-code:l1-start} -~\ref{line:clang-code:l1-end} are the 
body of the first loop with iterator \texttt{i} stored in \texttt{r8d}.
Lines~\ref{line:clang-code:l2-start} -~\ref{line:clang-code:l2-end} contain 
the body of the second loop with iterator \texttt{k} stored in \texttt{r9d}.
Lines~\ref{line:clang-code:l3-start} -~\ref{line:clang-code:jmp-l3} comprise 
the body of the third loop with iterator \texttt{j} stored in \texttt{rcx}.
Clang is able to eliminate a \texttt{cmp} instruction in the inner loop by 
initializing \texttt{rcx} with $-N_J$, incrementing \texttt{rcx} on each 
iteration at line~\ref{line:clang-code:rcx+1}, and using \texttt{jne} to test 
the zero flag of the status register, which is set to 1 when \texttt{rcx} becomes 0.

Figure~\ref{fig:v8-code} shows x86-64 code JITed by Chrome for the \wasm compiled 
version of \texttt{matmul}.
This code has been modified slightly -- \texttt{nop}s in the generated code 
have been removed for presentation.
Function arguments are passed in the \texttt{rax}, \texttt{rcx}, and \texttt{rdx} 
registers, following Chrome's calling convention. 
At lines~\ref{line:v8-code:rax-init}--~\ref{line:v8-code:rcx-init}, the contents 
of registers \texttt{rax}, \texttt{rdx}, and \texttt{rcx} are stored on the stack, 
due to registers spills at lines~\ref{line:v8-code:spill1} -~\ref{line:v8-code:spill3}.
Lines~\ref{line:v8-code:l1-start}--\ref{line:v8-code:l1-end} are the body of the
first loop with iterator \texttt{i} stored in \texttt{edi}.
Lines~\ref{line:v8-code:l2-start}--\ref{line:v8-code:l2-end} contain the body of 
second loop with iterator \texttt{k} stored in \texttt{r11}.
Lines~\ref{line:v8-code:l3-start}--\ref{line:v8-code:jmp-l3} are the body 
of the third loop with iterator \texttt{j} stored in \texttt{eax}.
To avoid memory loads due to register spilling at lines~\ref{line:v8-code:spill1}--~\ref{line:v8-code:spill3} 
in the first iteration of the first loop, an extra jump is generated at 
line~\ref{line:v8-code:jmp-L1'}.
Similarly, extra jumps are generated for the second and third loops at 
line~\ref{line:v8-code:jmp-L2'} and line~\ref{line:v8-code:jmp-L3'}
respectively.

\subsection{Differences}
The native code JITed by Chrome has more instructions, suffers from 
increased register pressure, and has extra branches compared to 
Clang-generated native code.

\subsubsection{Increased Code Size} 
\label{sec:ex:larger-code-size}
The number of instructions in the code generated by Chrome (Figure~\ref{fig:v8-code}) 
is 53, including \texttt{nop}s, while clang generated code (Figure~\ref{fig:clang-code}) 
consists of only 28 instructions.
The poor instruction selection algorithm of Chrome is one of the 
reasons for increased code size.

Additionally, Chrome does not take advantage of all available memory 
addressing modes for x86 instructions.
In Figure~\ref{fig:clang-code} Clang uses the \texttt{add} instruction at 
line~\ref{line:clang-code:addl} with register addressing mode, loading 
from and writing to a memory address in the same operation.
Chrome on the other hand loads the address in \texttt{ecx}, adds the 
operand to \texttt{ecx}, finally storing \texttt{ecx} at the address, 
requiring 3 instructions rather than one on 
lines~\ref{line:v8-code:addl-start}$-$\ref{line:v8-code:addl-end}.

\subsubsection{Increased Register Pressure} 
\label{sec:ex:reg-pressure}
Code generated by Clang in Figure~\ref{fig:clang-code} does not generate any spills and uses only 10 registers.
On the other hand, the code generated by Chrome (Figure~\ref{fig:v8-code}) uses 13 general purpose registers -- 
all available registers (\texttt{r13} and \texttt{r10} are reserved by V8).
As described in Section~\ref{sec:ex:larger-code-size}, eschewing the use of the register addressing 
mode of the \texttt{add} instruction requires the use of a temporary register.
All of this register inefficiency compounds, introducing three register spills to the stack at lines 1--3. Values stored on the stack are loaded again into registers at lines~\ref{line:v8-code:spill1}--\ref{line:v8-code:spill3} and
line~\ref{line:v8-code:l2-start}.

\subsubsection{Extra Branches}
\label{sec:ex:extra-jumps}
Clang (Figure~\ref{fig:clang-code}) generates code with a single branch per loop by
inverting the loop counter (line~\ref{line:clang-code:rcx+1}).
In contrast, Chrome (Figure~\ref{fig:v8-code}) generates more straightforward code,
which requires a conditional jump at the start of the loop.
In addition, Chrome generates extra jumps to avoid memory loads due to 
register spills in the first iteration of a loop. For example, the jump at
 line~\ref{line:v8-code:jmp-L1'} avoids the spills
at lines~\ref{line:v8-code:spill1}--~\ref{line:v8-code:spill3}.
\section{Performance Analysis}
\label{sec:spec}

\begin{table}
  \footnotesize
  \begin{tabular}{l|l}
      \textbf{\texttt{perf} Event}                                                     & \textbf{Wasm Summary}       \\ \hline
      \texttt{all-loads-retired (r81d0)} (Figure~\ref{fig:all-loads-retired})          & Increased register \\ 
      \texttt{all-stores-retired (r82d0)} (Figure~\ref{fig:all-stores-retired})        & pressure           \\ \hline
      \texttt{branches-retired (r00c4)} (Figure~\ref{fig:branch-instructions-retired}) & More branch        \\ 
      \texttt{conditional-branches (r01c4)} (Figure~\ref{fig:conditional-branches})    & statements         \\ \hline
      \texttt{instructions-retired (r1c0)} (Figure~\ref{fig:instructions-retired})     & Increased code size     \\ 
      \texttt{cpu-cycles} (Figure~\ref{fig:cpu-cycles})                                &                      \\ 
      \texttt{L1-icache-load-misses} (Figure~\ref{fig:icache-load-misses})             &                    \\ 
  \end{tabular}
  \caption{
    Performance counters highlight
    specific issues with \wasm{} code generation. When a raw PMU
    event descriptor is used, it is indicated by $\texttt{r}n$.}
  \label{table:perf-events}
\end{table}

\begin{figure*}[!t]

\begin{subfigure}[b]{0.33\textwidth}
\begin{tikzpicture}
\node{\pgfimage[width=\textwidth]{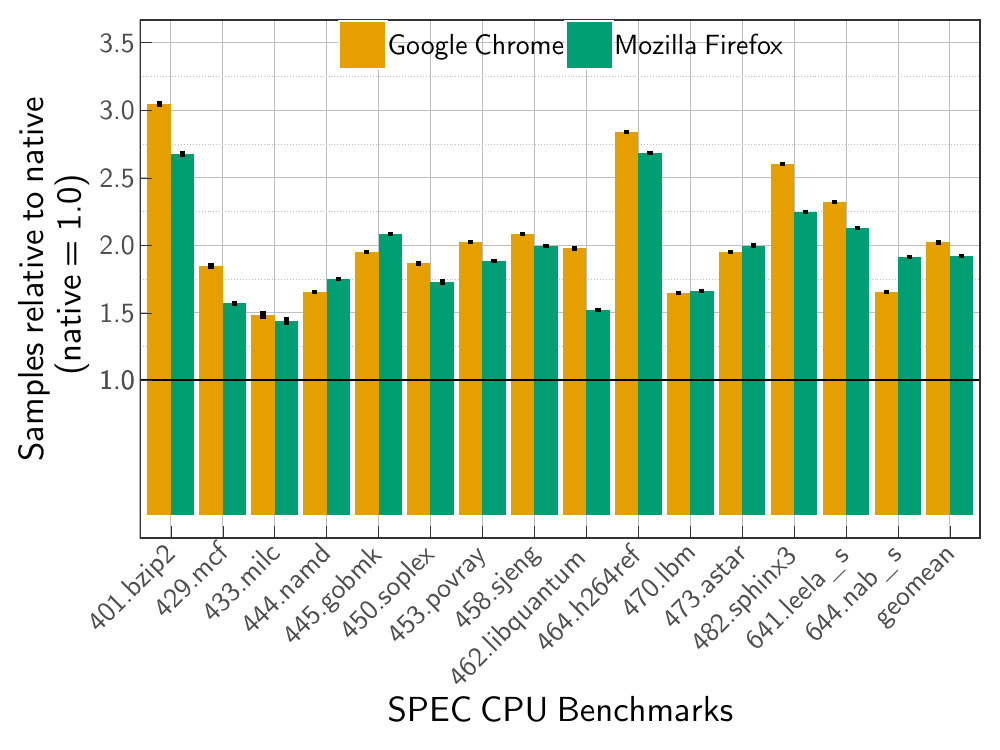}};
\end{tikzpicture}
\caption{\texttt{all-loads-retired}}
\label{fig:all-loads-retired}
\end{subfigure}
\begin{subfigure}[b]{0.33\textwidth}
\begin{tikzpicture}
\node{\pgfimage[width=\textwidth]{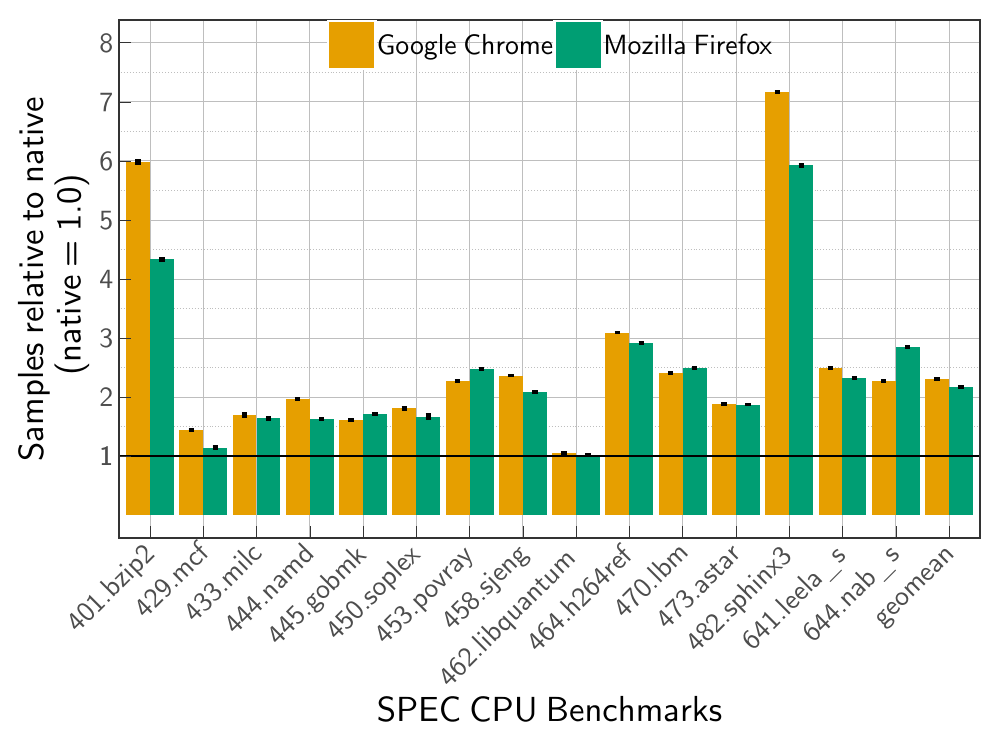}};
\end{tikzpicture}
\caption{\texttt{all-stores-retired}}
\label{fig:all-stores-retired}
\end{subfigure}
\begin{subfigure}[b]{0.33\textwidth}
\begin{tikzpicture}
\node{\pgfimage[width=\textwidth]{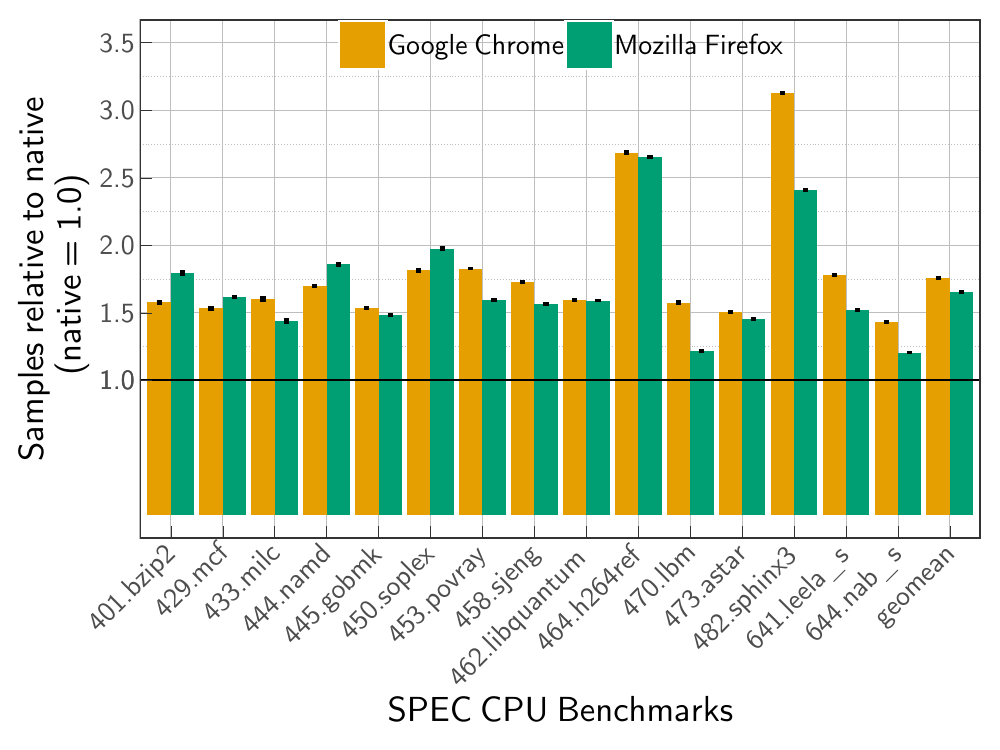}};
\end{tikzpicture}
\caption{\texttt{branch-instructions-retired}\label{fig:branch-instructions-retired}}
\end{subfigure}

\begin{subfigure}[b]{0.33\textwidth}
\begin{tikzpicture}
\node{\pgfimage[width=\textwidth]{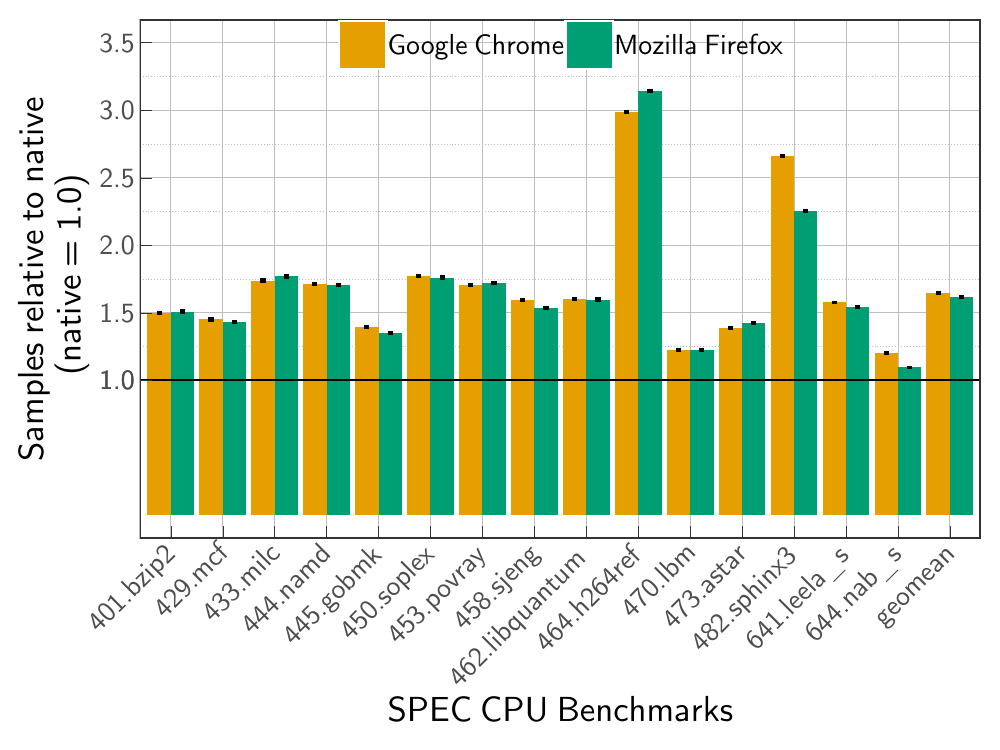}};
\end{tikzpicture}
\caption{\texttt{conditional-branches}\label{fig:conditional-branches}}
\end{subfigure}
\begin{subfigure}[b]{0.33\textwidth}
\begin{tikzpicture}
\node{\pgfimage[width=\textwidth]{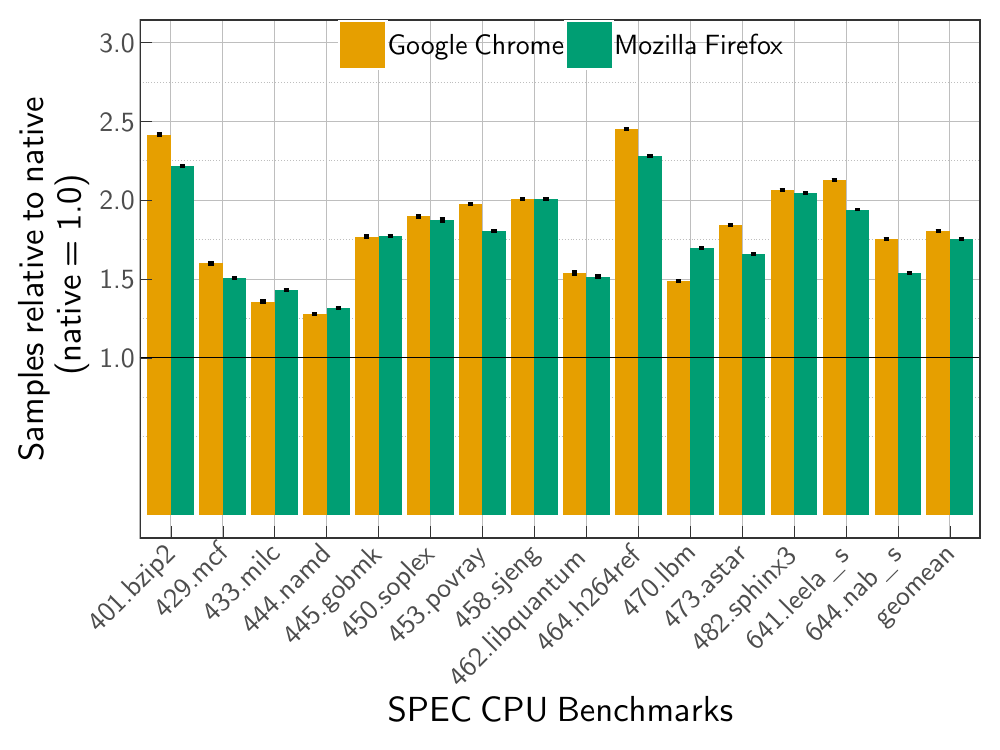}};
\end{tikzpicture}
\caption{\texttt{instructions-retired}}
\label{fig:instructions-retired}
\end{subfigure}
\begin{subfigure}[b]{0.33\textwidth}
\begin{tikzpicture}
\node{\pgfimage[width=\textwidth]{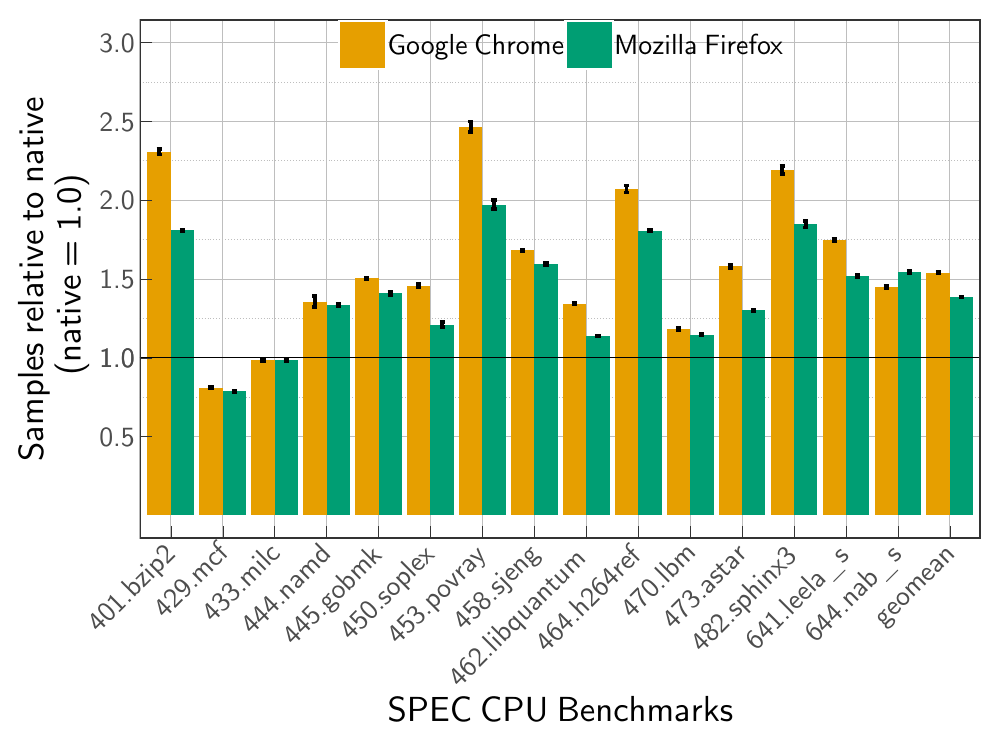}};
\end{tikzpicture}
\caption{\texttt{cpu-cycles}}
\label{fig:cpu-cycles}
\end{subfigure}
\caption{Processor performance counter samples for \wasm{} relative to
native code.}

\end{figure*}

We use \BrowsixSPEC to record measurements from all supported performance
counters on our system for the SPEC CPU benchmarks compiled
to \wasm{} and executed in Chrome and Firefox, and
the SPEC CPU benchmarks compiled to native code 
(Section~\ref{sec:system:bench-setup}).

Table~\ref{table:perf-events} lists the performance counters we use
here, along with a summary of the impact of \BrowsixWasm performance
on these counters compared to native.  We use these results to explain
the performance overhead of \wasm{} over native code. Our analysis
shows that the inefficiences described in Section~\ref{sec:ex}
are pervasive and translate to reduced performance across the SPEC CPU
benchmark suite.

\subsection{Increased Register Pressure}
\label{sec:spec:increased-reg-pressure}

This section focuses on two performance counters that show the effect
of increased register pressure.  Figure~\ref{fig:all-loads-retired}
presents the number of \emph{load} instructions retired by \wasm{}-compiled
SPEC benchmarks in Chrome and Firefox, relative to the number of load
instructions retired in native code. Similarly,
Figure~\ref{fig:all-stores-retired} shows the number of \emph{store}
instructions retired.  Note that a ``retired'' instruction is an
instruction which leaves the instruction pipeline and its results are
correct and visible in the architectural state (that is, not
speculative).

Code generated by Chrome has 2.02$\times$ more load
instructions retired and 2.30$\times$ more store instructions retired than native code. 
Code generated by Firefox has 1.92$\times$ more load instructions
retired and 2.16$\times$ more store instructions retired than native
code.
These results show that the \wasm{}-compiled SPEC CPU
benchmarks suffer from increased register pressure and thus
increased memory references.  Below, we outline the reasons
for this increased register pressure.

\subsubsection{Reserved Registers}
\label{sec:spec:less-regs}

In Chrome,
\texttt{matmul} generates three register spills but does not use two
x86-64 registers: \texttt{r13} and \texttt{r10}
(Figure~\ref{fig:v8-code}, lines~\ref{line:v8-code:spill1}--~\ref{line:v8-code:spill3}).
This occurs because Chrome reserves these two registers.\footnote{\url{https://github.com/v8/v8/blob/7.4.1/src/x64/register-x64.h}}
For the JavaScript garbage collector, Chrome reserves \texttt{r13} to
point to an array of GC roots at all times. In addition, Chrome
uses \texttt{r10} and \texttt{xmm13} as dedicated scratch registers.
Similarly, Firefox reserves \texttt{r15} as a pointer to the start of the heap,
and \texttt{r11} and \texttt{xmm15} are JavaScript scratch registers.\footnote{\url{https://hg.mozilla.org/mozilla-central/file/tip/js/src/jit/x64/Assembler-x64.h}}
None of these registers are available to \wasm{} code.

\subsubsection{Poor Register Allocation}
\label{sec:reg-allocator}

Beyond a reduced set of registers available to allocate, both Chrome
and Firefox do a poor job of allocating the registers they have.  For
example, the code generated by Chrome for \texttt{matmul} uses
12 registers while the native code generated by Clang only uses 10
registers (Section~\ref{sec:ex:reg-pressure}). This increased register
usage---in both Firefox and Chrome---is because of their use of fast
but not particularly effective register allocators. Chrome and Firefox
both use a linear scan register
allocator~\cite{wimmer-linear-scan-ssa}, while Clang uses a
greedy graph-coloring register allocator~\cite{llvm-ref-manual}, which consistently generates better code.

\subsubsection{x86 Addressing Modes}
The x86-64 instruction set offers several addressing modes for each
operand, including a \emph{register} mode, where the instruction reads
data from register or writes data to a register, and memory address
modes like \emph{register indirect} or \emph{direct offset}
addressing, where the operand resides in a memory address and the
instruction can read from or write to that address. A code generator
could avoid unnecessary register pressure by using the latter
modes. However, Chrome does not take advantage of these modes.  For
example, the code generated by Chrome for \texttt{matmul} does not use
the register indirect addressing mode for the \texttt{add} instruction
(Section~\ref{sec:ex:reg-pressure}), creating unnecessary register
pressure.

\subsection{Extra Branch Instructions}
\label{sec:spec:extra-branch}

This section focuses on two performance counters that measure the number of branch instructions
executed.
Figure~\ref{fig:branch-instructions-retired} shows the number of branch 
instructions retired by \wasm{},
relative to the number of branch instructions
retired in native code. Similarly, Figure~\ref{fig:conditional-branches} 
shows the number of \emph{conditional} branch instructions retired.
In Chrome, there are $1.75\times$ and $1.65\times$ more unconditional
and conditional branch instructions retired respectively, whereas in Firefox, there
are  $1.65\times$ and  $1.62\times$ more retired.
These results show that all the SPEC CPU benchmarks incur extra branches, and
we explain why below.

\subsubsection{Extra Jump Statements for Loops}
\label{extra-jumps}

As with \texttt{matmul} (Section~\ref{sec:ex:extra-jumps}), Chrome
generates unnecessary jump statements for loops, leading to 
significantly more branch instructions than Firefox.

\subsubsection{Stack Overflow Checks Per Function Call}
\label{stack-overflow}

A \wasm{} program tracks the current stack size with a global variable
that it increases on every function call. The programmer can define
the maximum stack size for the program.  To ensure that a program does
not overflow the stack, both Chrome and Firefox add stack checks at
the start of each function to detect if the current stack size is less
than the maximum stack size. These checks includes extra comparison
and conditional jump instructions, which must be executed on every
function call.

\subsubsection{Function Table Indexing Checks}
\label{indirect-call-checks}

\wasm{} dynamically checks all indirect calls to ensure that the target is a
valid function and that the function's type at runtime is the same as the type
specified at the call site. In a \wasm{} module, the function table stores
the list of functions and their types, and the code generated by \wasm{}
uses the function table to implement these checks.
These checks are required when calling function
pointers and virtual functions in C/C++. The checks lead to extra comparison
and conditional jump instructions, which are executed before every indirect
function call.

\subsection {Increased Code Size}
\label{increased-code-size}

\begin{figure}[t]
\centering
\begin{tikzpicture}
\node{\pgfimage[width=1.0\columnwidth]{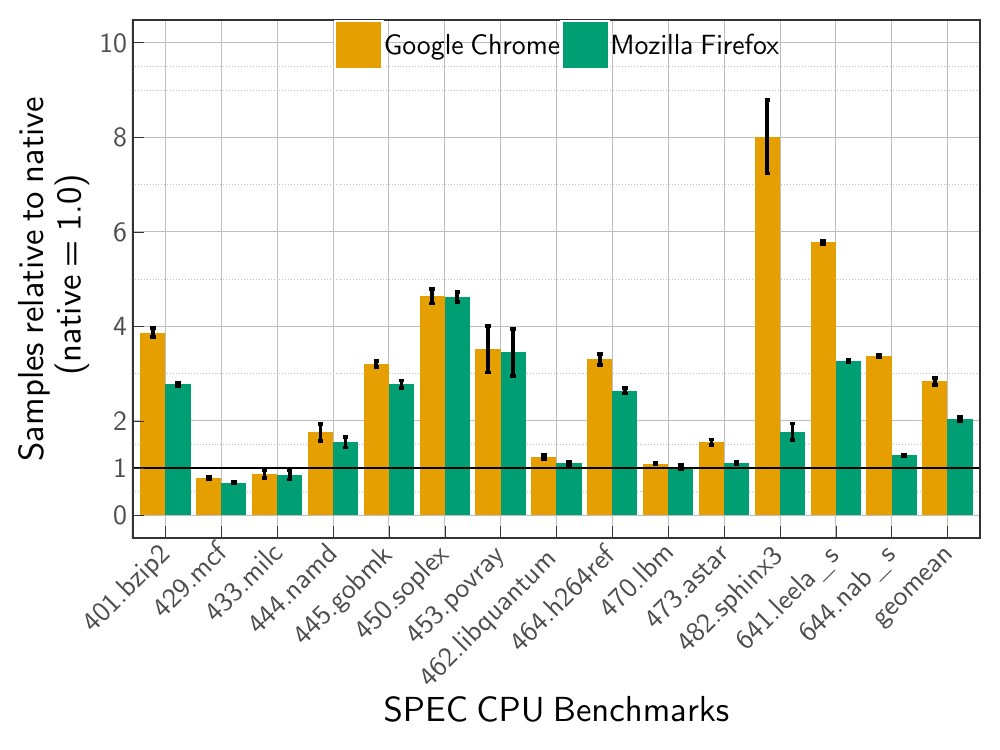}};
\end{tikzpicture}
\caption{ \texttt{L1-icache-load-misses} samples counted for SPEC
    CPU compiled to \wasm{} executed in Chrome and Firefox, relative
    to native. \texttt{458.sjeng} not shown in the graph exhibits 26.5$\times$ more L1 instruction cache misses in Chrome and 18.6$\times$ more in Firefox.
    The increased code size generated for \wasm{} leads to
    more instruction cache misses.
    \label{fig:icache-load-misses}}
\end{figure}

 The  code generated by Chrome and Firefox is considerably larger than
 the code generated by Clang.
We use three performance counters to measure this effect.  (i)
Figure~\ref{fig:instructions-retired} shows the number of
\emph{instructions retired} by benchmarks compiled to \wasm{} and
executed in Chrome and Firefox relative to the number of instructions
retired in native code.  Similarly, Figure~\ref{fig:cpu-cycles} shows
the relative number of \emph{CPU cycles} spent by benchmarks compiled to
\wasm{}, and Figure~\ref{fig:icache-load-misses} shows the relative number of
\emph{L1 instruction cache load misses}.

Figure~\ref{fig:instructions-retired} shows that Chrome executes
an average of 1.80$\times$ more instructions than native code and Firefox executes an average of 1.75$\times$ more
instructions than native code. Due to poor instruction
selection, a poor register allocator generating more register spills
(Section~\ref{sec:spec:increased-reg-pressure}), and extra branch
statements (Section~\ref{sec:spec:extra-branch}), the size of generated
code for \wasm{} is greater than native code, leading to more
instructions being executed. This increase in the number of
instructions executed leads to increased L1 instruction cache misses
in Figure~\ref{fig:icache-load-misses}.  On average, Chrome suffers
2.83$\times$ more I-cache misses than native code,
and Firefox suffers from 2.04$\times$ more L1 instruction cache misses
than native code. More cache misses means that more CPU cycles are
spent waiting for the instruction to be fetched.

We note one anomaly: although \texttt{429.mcf} has 1.6$\times$ more
instructions retired in Chrome than native code and 1.5$\times$ more
instructions retired in Firefox than native code, it runs \emph{faster}
than native code. Figure~\ref{fig:spec-relative-times} shows that its
slowdown relative to native is 0.81$\times$ in Chrome and 0.83$\times$ in Firefox.
The reason for this anomaly is
attributable directly to its lower number of L1 instruction cache
misses. \texttt{429.mcf} contains a main loop and most of the instructions in the loop fit in the L1 instruction cache. Similarly,
\texttt{433.milc} performance is better due to fewer L1 instruction cache misses. In \texttt{450.soplex} there are 4.6$\times$ more L1 instruction cache misses in Chrome and Firefox than native because of several virtual functions being executed, leading to more indirect function calls. 

\begin{table}
  \begin{tabular}{l|c|c}
    \hline
    \textbf{Performance Counter}                  & \textbf{Chrome} & \textbf{Firefox} \\ \hline
    \texttt{all-loads-retired}           & 2.02$\times$   & 1.92$\times$ \\ 
    \texttt{all-stores-retired}          & 2.30$\times$     & 2.16$\times$ \\ 
    \texttt{branch-instructions-retired} & 1.75$\times$  & 1.65$\times$ \\
    \texttt{conditional-branches}        & 1.65$\times$  & 1.62$\times$ \\ 
    \texttt{instructions-retired}        & 1.80$\times$   & 1.75$\times$ \\ 
    \texttt{cpu-cycles}                  & 1.54$\times$  & 1.38$\times$ \\ 
    \texttt{L1-icache-load-misses}       & 2.83$\times$  & 2.04$\times$ \\
    \hline
  \end{tabular}
  \caption {The geomean of performance counter increases for the SPEC benchmarks using \wasm{}.\label{tab:geomean-perf}}
  \vspace{-1em}
\end{table}

\subsection{Discussion}

It is worth asking if the performance issues identified here are
fundamental. We believe that two of the identified issues are not:
that is, they could be ameliorated by improved implementations.
\wasm{} implementations today use register allocators
(\S\ref{sec:reg-allocator}) and code generators (\S\ref{extra-jumps})
that perform worse than Clang's counterparts. However, an offline
compiler like Clang can spend considerably more time to generate
better code, whereas \wasm{} compilers must be fast enough to run
online. Therefore, solutions adopted by other JITs, such as
further optimizing hot code, are likely applicable
here~\cite{hotspot-client, adaptive-jit}.

The four other issues that we have identified appear to arise from the
design constraints of \wasm{}: the stack
overflow checks (\S\ref{stack-overflow}), indirect call checks
(\S\ref{indirect-call-checks}), and reserved registers (\S\ref{sec:spec:less-regs})
have  a runtime cost and lead to increased code
size (\S\ref{increased-code-size}). Unfortunately, these checks are
necessary for \wasm{}'s safety guarantees. A redesigned \wasm{}, with richer
types for memory and function pointers~\cite{tal}, might be able to perform
some of these checks at compile time, but that could complicate
the implementation of compilers that produce \wasm{}.
Finally, a \wasm{} implementation in a browser must interoperate with a
high-performance JavaScript implementation, which may impose its own
constraints. For example, current JavaScript implementations reserve
a few registers for their own use, which increases register pressure
on \wasm{}.

\section{Related Work}

\paragraph{Precursors to WebAssembly}

There have been several attempts to execute native code in browsers, but none
of them met all the design criteria of \wasm{}.

ActiveX~\cite{activeX} allows web pages to embed signed x86 libraries,
however these binaries have unrestricted access to the Windows API.
In contrast, \wasm{} modules are sandboxed. ActiveX is now a deprecated
technology.

Native Client~\cite{yee-nacl, ansel-sandboxing} (NaCl)
adds a module to a web application that contains platform specific 
machine code. NaCl introduced sandboxing techniques to execute platform 
specific machine code at near native speed. Since NaCl relies on static 
validation of machine code, it requires code generators to follow certain
patterns, hence, supporting only a subset of the x86, ARM, and MIPS instructions sets in 
the browser. To address the inherent portability issue of NaCl, Portable NaCl 
(PNaCl)~\cite{pnacl} uses LLVM Bitcode as a binary format. However, PNaCl does not 
provide significant improvement in compactness over NaCl and still exposes 
compiler and/or platform-specific details such as the call stack layout. Both 
have been deprecated in favor of \wasm{}. 

\asmjs{} is a subset of JavaScript 
designed to be compiled efficiently to native code. \asmjs{} uses type coercions 
to avoid the dynamic type system of JavaScript. Since \asmjs{} is a subset of JavaScript,
adding all native features to \asmjs{} such as 64-bit integers will first require extending
JavaScript. Compared to \asmjs{}, \wasm{} provides several improvements: 
~(i) \wasm{} binaries are compact due to its lightweight representation compared to 
JavaScript source, (ii) \wasm{} is more straightforward to validate, 
(iii) \wasm{} provides formal guarantees of type safety and isolation, and
(iv) \wasm{} has been shown to provide better performance than \asmjs {}.

\wasm{} is a stack machine, which is similar to the Java Virtual
Machine~\cite{jvm} and the Common Language Runtime~\cite{cil}. However, \wasm{}
is very different from these platforms. For example \wasm{} does not support objects
and does not support unstructured control flow.

The \wasm{} specification defines its operational semantics and type system.
This proof was mechanized using the
Isabelle theorem prover, and that mechanization effort found and addressed a
number of issues in the specification~\cite{watt:wasm-verif}.
RockSalt~\cite{morrisett:rocksalt} is a similar verification effort for NaCl.
It implements the NaCl verification toolchain in Coq, along with a proof of
correctness with respect to a model of the subset of x86 instructions that
NaCl supports.

\paragraph{Analysis of SPEC Benchmarks using performance counters}
Several papers use performance counters to analyze the SPEC
benchmarks. Panda et al.~\cite{panda-wait-of-a-decade} analyze the
SPEC CPU2017 benchmarks, applying statistical techniques to identify similarities
among benchmarks. Phansalkar et al. perform a similar study on SPEC
CPU2006~\cite{phansalkar-spec2006-analysis}. Limaye and Adegija
 identify workload differences between
SPEC CPU2006 and SPEC CPU2017~\cite{limaye-spec2017-workload}.

\section{Conclusions}

This paper performs the first comprehensive performance analysis of
WebAssembly. We develop \browsixwasm, a significant extension of
\Browsix, and \BrowsixSpec, a harness that enables detailed
performance analysis, to let us run the SPEC CPU2006 and CPU2017
benchmarks as \wasm in Chrome and Firefox. We find that the mean
slowdown of \wasm{} vs. native across SPEC benchmarks is 1.55$\times$ for Chrome and 1.45$\times$
for Firefox, with peak slowdowns of
2.5$\times$ in Chrome and 2.08$\times$ in Firefox. We identify the
causes of these performance gaps, providing actionable guidance for
future optimization efforts.

\vspace{-1em}
\paragraph{Acknowledgements} We thank the reviewers and our shepherd, Eric
Eide, for their constructive feedback. This work was partially supported
by NSF grants 1439008 and 1413985.

\clearpage
\bibliographystyle{plain}
\interlinepenalty=10000
\bibliography{emery,main}

\begin{thebibliography}{10}

\bibitem{blazor}
{Blazor}.
\newblock \url{https://blazor.net/}.
\newblock [Online; accessed 5-January-2019].

\bibitem{rust-wasm}
Compiling from {Rust} to {WebAssembly}.
\newblock
  \url{https://developer.mozilla.org/en-US/docs/WebAssembly/Rust_to_wasm}.
\newblock [Online; accessed 5-January-2019].

\bibitem{llvm-ref-manual}
{LLVM Reference Manual}.
\newblock \url{https://llvm.org/docs/CodeGenerator.html}.

\bibitem{nacl-and-pnacl}
{NaCl} and {PNaCl}.
\newblock \url{https://developer.chrome.com/native-client/nacl-and-pnacl}.
\newblock [Online; accessed 5-January-2019].

\bibitem{polybench}
{PolyBenchC}: the polyhedral benchmark suite.
\newblock \url{http://web.cs.ucla.edu/~pouchet/software/polybench/}.
\newblock [Online; accessed 14-March-2017].

\bibitem{js-heap-limit}
{Raise Chrome JS heap limit? - Stack Overflow}.
\newblock
  \url{https://stackoverflow.com/questions/43643406/raise-chrome-js-heap-limit}.
\newblock [Online; accessed 5-January-2019].

\bibitem{wasm-use-cases}
Use cases.
\newblock \url{https://webassembly.org/docs/use-cases/}.

\bibitem{webassembly}
{WebAssembly}.
\newblock \url{https://webassembly.org/}.
\newblock [Online; accessed 5-January-2019].

\bibitem{linux-calling-conventions}
{System V Application Binary Interface AMD64 Architecture Processor
  Supplement}.
\newblock
  \url{https://software.intel.com/sites/default/files/article/402129/mpx-linux64-abi.pdf},
  2013.

\bibitem{curated-wasm-list}
Steve Akinyemi.
\newblock {A curated list of languages that compile directly to or have their
  {VMs} in {WebAssembly}}.
\newblock \url{https://github.com/appcypher/awesome-wasm-langs}.
\newblock [Online; accessed 5-January-2019].

\bibitem{ansel-sandboxing}
Jason Ansel, Petr Marchenko, \'{U}lfar Erlingsson, Elijah Taylor, Brad Chen,
  Derek~L. Schuff, David Sehr, Cliff~L. Biffle, and Bennet Yee.
\newblock {Language-independent Sandboxing of Just-in-time Compilation and
  Self-modifying Code}.
\newblock In {\em {Proceedings of the 32nd ACM SIGPLAN Conference on
  Programming Language Design and Implementation}}, PLDI '11, pages 355--366.
  ACM, 2011.

\bibitem{bebenita:spur}
Michael Bebenita, Florian Brandner, Manuel Fahndrich, Francesco Logozzo,
  Wolfram Schulte, Nikolai Tillmann, and Herman Venter.
\newblock {SPUR: A Trace-based JIT Compiler for CIL}.
\newblock In {\em {Proceedings of the ACM International Conference on Object
  Oriented Programming Systems Languages and Applications}}, OOPSLA '10, pages
  708--725. ACM, 2010.

\bibitem{activeX}
David~A Chappell.
\newblock {\em Understanding {ActiveX} and {OLE}}.
\newblock Microsoft Press, 1996.

\bibitem{pnacl}
Alan Donovan, Robert Muth, Brad Chen, and David Sehr.
\newblock {{PNaCl}: Portable Native Client Executables}.
\newblock \url{https://css.csail.mit.edu/6.858/2012/readings/pnacl.pdf}, 2010.

\bibitem{eich:2015wasm}
Brendan Eich.
\newblock From {ASM.JS} to {WebAssembly}.
\newblock \url{https://brendaneich.com/2015/06/from-asm-js-to-webassembly/},
  2015.
\newblock [Online; accessed 5-January-2019].

\bibitem{elliott:2015wasm}
Eric Elliott.
\newblock What is {WebAssembly}?
\newblock \url{https://tinyurl.com/o5h6daj}, 2015.
\newblock [Online; accessed 5-January-2019].

\bibitem{gal:tracemonkey}
Andreas Gal, Brendan Eich, Mike Shaver, David Anderson, David Mandelin,
  Mohammad~R. Haghighat, Blake Kaplan, Graydon Hoare, Boris Zbarsky, Jason
  Orendorff, Jesse Ruderman, Edwin~W. Smith, Rick Reitmaier, Michael Bebenita,
  Mason Chang, and Michael Franz.
\newblock {Trace-based Just-in-time Type Specialization for Dynamic Languages}.
\newblock In {\em {Proceedings of the 30th ACM SIGPLAN Conference on
  Programming Language Design and Implementation}}, PLDI '09, pages 465--478.
  ACM, 2009.

\bibitem{haas:2017:webassembly}
Andreas Haas, Andreas Rossberg, Derek~L. Schuff, Ben~L. Titzer, Michael Holman,
  Dan Gohman, Luke Wagner, Alon Zakai, and JF~Bastien.
\newblock {Bringing the Web Up to Speed with WebAssembly}.
\newblock In {\em {Proceedings of the 38th ACM SIGPLAN Conference on
  Programming Language Design and Implementation}}, PLDI 2017, pages 185--200.
  ACM, 2017.

\bibitem{hotspot-client}
Thomas Kotzmann, Christian Wimmer, Hanspeter M\"{o}ssenb\"{o}ck, Thomas
  Rodriguez, Kenneth Russell, and David Cox.
\newblock {Design of the Java HotSpot Client Compiler for Java 6}.
\newblock {\em {ACM Trans. Archit. Code Optim.}}, 5(1):7:1--7:32, 2008.

\bibitem{limaye-spec2017-workload}
Ankur Limaye and Tosiron Adegbija.
\newblock {A Workload Characterization of the SPEC CPU2017 Benchmark Suite}.
\newblock In {\em {2018 IEEE International Symposium on Performance Analysis of
  Systems and Software (ISPASS)}}, pages 149--158, 2018.

\bibitem{jvm}
Tim Lindholm, Frank Yellin, Gilad Bracha, and Alex Buckley.
\newblock {\em The Java Virtual Machine Specification, Java SE 8 Edition}.
\newblock Addison-Wesley Professional, 1st edition, 2014.

\bibitem{morrisett:rocksalt}
Greg Morrisett, Gang Tan, Joseph Tassarotti, Jean-Baptiste Tristan, and Edward
  Gan.
\newblock {RockSalt: Better, Faster, Stronger SFI for the x86}.
\newblock In {\em {Proceedings of the 33rd ACM SIGPLAN Conference on
  Programming Language Design and Implementation}}, PLDI '12, pages 395--404.
  ACM, 2012.

\bibitem{tal}
Greg Morrisett, David Walker, Karl Crary, and Neal Glew.
\newblock {From System F to Typed Assembly Language}.
\newblock {\em {ACM Trans. Program. Lang. Syst.}}, 21(3):527--568, 1999.

\bibitem{musiol:2016gopherjs}
Richard Musiol.
\newblock A compiler from {Go} to {JavaScript} for running {Go} code in a
  browser.
\newblock \url{https://github.com/gopherjs/gopherjs}, 2016.
\newblock [Online; accessed 5-January-2019].

\bibitem{cil}
George~C. Necula, Scott McPeak, Shree~P. Rahul, and Westley Weimer.
\newblock {CIL: Intermediate Language and Tools for Analysis and Transformation
  of C Programs}.
\newblock In R.~Nigel Horspool, editor, {\em {Compiler Construction}}, pages
  213--228. Springer, 2002.

\bibitem{panda-wait-of-a-decade}
Reena Panda, Shuang Song, Joseph Dean, and Lizy~K. John.
\newblock {Wait of a Decade: Did SPEC CPU 2017 Broaden the Performance
  Horizon?}
\newblock In {\em {2018 IEEE International Symposium on High Performance
  Computer Architecture (HPCA)}}, pages 271--282, 2018.

\bibitem{phansalkar-spec2006-analysis}
Aashish Phansalkar, Ajay Joshi, and Lizy~K. John.
\newblock {Analysis of Redundancy and Application Balance in the SPEC CPU2006
  Benchmark Suite}.
\newblock In {\em {Proceedings of the 34th Annual International Symposium on
  Computer Architecture}}, ISCA '07, pages 412--423. ACM, 2007.

\bibitem{browsix-web}
Bobby Powers, John Vilk, and Emery~D. Berger.
\newblock Browsix: {Unix} in your browser tab.
\newblock \url{https://browsix.org}.

\bibitem{powers:2017:browsix}
Bobby Powers, John Vilk, and Emery~D. Berger.
\newblock {Browsix: Bridging the Gap Between {Unix} and the Browser}.
\newblock In {\em {Proceedings of the Twenty-Second International Conference on
  Architectural Support for Programming Languages and Operating Systems}},
  ASPLOS '17, pages 253--266. ACM, 2017.

\bibitem{richards:js-behavior}
Gregor Richards, Sylvain Lebresne, Brian Burg, and Jan Vitek.
\newblock {An Analysis of the Dynamic Behavior of JavaScript Programs}.
\newblock In {\em {Proceedings of the 31st ACM SIGPLAN Conference on
  Programming Language Design and Implementation}}, PLDI '10, pages 1--12. ACM,
  2010.

\bibitem{selakovic:js-optimizations}
Marija Selakovic and Michael Pradel.
\newblock {Performance Issues and Optimizations in JavaScript: An Empirical
  Study}.
\newblock In {\em {Proceedings of the 38th International Conference on Software
  Engineering}}, ICSE '16, pages 61--72. ACM, 2016.

\bibitem{adaptive-jit}
Toshio Suganuma, Toshiaki Yasue, Motohiro Kawahito, Hideaki Komatsu, and Toshio
  Nakatani.
\newblock {A Dynamic Optimization Framework for a Java Just-in-time Compiler}.
\newblock In {\em {Proceedings of the 16th ACM SIGPLAN Conference on
  Object-oriented Programming, Systems, Languages, and Applications}}, OOPSLA
  '01, pages 180--195. ACM, 2001.

\bibitem{asmjs:announcement}
Luke Wagner.
\newblock {asm.js in Firefox Nightly | Luke Wagner's Blog}.
\newblock
  \url{https://blog.mozilla.org/luke/2013/03/21/asm-js-in-firefox-nightly/}.
\newblock [Online; accessed 21-May-2019].

\bibitem{wagner:2016wasm}
Luke Wagner.
\newblock {A WebAssembly Milestone: Experimental Support in Multiple Browsers}.
\newblock \url{https://hacks.mozilla.org/2016/03/a-webassembly-milestone/},
  2016.
\newblock [Online; accessed 5-January-2019].

\bibitem{watt:wasm-verif}
Conrad Watt.
\newblock {Mechanising and Verifying the WebAssembly Specification}.
\newblock In {\em {Proceedings of the 7th ACM SIGPLAN International Conference
  on Certified Programs and Proofs}}, CPP 2018, pages 53--65. ACM, 2018.

\bibitem{wimmer-linear-scan-ssa}
Christian Wimmer and Michael Franz.
\newblock {Linear Scan Register Allocation on SSA Form}.
\newblock In {\em {Proceedings of the 8th Annual IEEE/ACM International
  Symposium on Code Generation and Optimization}}, CGO '10, pages 170--179.
  ACM, 2010.

\bibitem{yee-nacl}
Bennet Yee, David Sehr, Greg Dardyk, Brad Chen, Robert Muth, Tavis Ormandy,
  Shiki Okasaka, Neha Narula, and Nicholas Fullagar.
\newblock {Native Client: A Sandbox for Portable, Untrusted x86 Native Code}.
\newblock In {\em {IEEE Symposium on Security and Privacy (Oakland'09)}}, IEEE,
  2009.

\bibitem{asmjs}
Alon Zakai.
\newblock asm.js.
\newblock \url{http://asmjs.org/}.
\newblock [Online; accessed 5-January-2019].

\bibitem{emscripten}
Alon Zakai.
\newblock {Emscripten: An {LLVM-to-JavaScript} Compiler}.
\newblock In {\em {Proceedings of the ACM International Conference Companion on
  Object Oriented Programming Systems Languages and Applications Companion}},
  OOPSLA '11, pages 301--312. ACM, 2011.

\end{thebibliography}

\end{document}